\documentclass[aps,prb,twocolumn,showpacs,preprintnumbers]{revtex4}
\usepackage{graphicx} %Include figure files
\usepackage{amsmath}
\newcommand{\sisj}{\langle\mathbf S_i\cdot\mathbf S_j\rangle}
\newcommand{\qv}{\mathbf q}

\newcommand{\be}{\begin{equation}}
\newcommand{\ee}{\end{equation}}
\newcommand{\bea}{\begin{eqnarray}}
\newcommand{\eea}{\end{eqnarray}}
\newcommand{\bse}{\begin{subequations}}
\newcommand{\ese}{\end{subequations}}

\newcommand{\bi}{$\rm BiMn_2PO_6$}

\begin{document}

\title{Magnetic transitions in the spin ${\bf S=\frac52}$ frustrated magnet BiMn$_{2}$PO$_{6}$\\ and strong lattice softening in BiMn$_{2}$PO$_{6}$ and BiZn$_{2}$PO$_{6}$ below 200~K}

\author{R. Nath}
\email{rnath@iisertvm.ac.in}
\affiliation{School of Physics, Indian Institute of Science Education and
Research, Thiruvananthapuram-695016, Kerala, India}
\author{B. Roy}
\affiliation{Ames Laboratory and Department of Physics and Astronomy, Iowa
State University, Ames, IA 50011, USA}
\author{K. M. Ranjith}
\affiliation{School of Physics, Indian Institute of Science Education and
Research, Thiruvananthapuram-695016, Kerala, India}
\author{D. C. Johnston}
\author{Y. Furukawa}
\affiliation{Ames Laboratory and Department of Physics and Astronomy, Iowa
State University, Ames, IA 50011, USA}

\author{A. A. Tsirlin}
\email{altsirlin@gmail.com}
\affiliation{National Institute of Chemical Physics and Biophysics, 12638 Tallinn, Estonia}

\date{\today}

\begin{abstract}

The crystallographic, magnetic and thermal properties of polycrystalline BiMn$_{2}$PO$_{6}$ and its nonmagnetic analogue BiZn$_{2}$PO$_{6}$ were investigated by x-ray diffraction, magnetization~$M$, magnetic susceptibility~$\chi$, heat capacity $C_{\rm p}$, and $^{31}$P nuclear magnetic resonance (NMR) measurements versus applied magnetic field~$H$ and temperature~$T$ as well as by density-functional band-theory and molecular field calculations.  Both compounds show a strong monotonic lattice softening on cooling, where the Debye temperature decreases by a factor of two from $\Theta_{\rm D}\sim 650$~K at $T = 300$~K to $\Theta_{\rm D}\sim 300$~K at $T=2$~K\@.  The $\chi(T)$ data for BiMn$_{2}$PO$_{6}$ above 150~K follow a Curie-Weiss law with a Curie constant consistent with a Mn$^{+2}$ spin~$S = 5/2$ with $g$-factor $g=2$ and an antiferromagnetic (AFM) Weiss temperature $\theta_{\rm CW}\simeq -78$\,K\@.  The $\chi$ data indicate long-range AFM ordering below $T_{\rm N}\simeq 30$~K, confirmed by a sharp $\lambda$-shaped peak in $C_{\rm p}(T)$ at 28.8~K\@.  The magnetic entropy at 100~K extracted from the $C_{\rm p}(T)$ data is consistent with spin $S = 5/2$ for the Mn$^{+2}$ cations.  The band-theory calculations indicate that BiMn$_2$PO$_6$ is an AFM compound with dominant interactions $J_1/k_{\rm B}\simeq 6.7$\,K and $J_3/k_{\rm B}\simeq 5.6$\,K along the legs and rungs of a Mn two-leg spin-ladder, respectively. However, sizable and partially frustrating interladder couplings lead to an anisotropic three-dimensional magnetic behavior with long-range AFM ordering at $T_{\rm N}\simeq 30$\,K observed in the $\chi$, $C_{\rm p}$ and NMR measurements. A second magnetic transition at $\approx10$\,K is observed from the $\chi$ and NMR measurements but is not evident in the $C_{\rm p}$ data. The $C_{\rm p}$ data at low~$T$ suggest a significant contribution from AFM spin waves moving in three dimensions and the absence of a spin-wave gap.  A detailed analysis of the NMR spectra indicates commensurate magnetic order between 10\,K and 30\,K, while below 10\,K additional features appear that may arise from an incommensurate modulation and/or spin canting.  The commensurate order is consistent with microscopic density functional calculations that yield a collinear N\'eel-type  AFM spin arrangement both within and between the ladders, despite the presence of multiple weak interactions frustrating this magnetic structure of the Mn spins. Frustration for AFM ordering and the 1D spatial anisotropy of the 3D spin interactions are manifested in the frustration ratio $f = |\theta_{\rm CW}|/T_{\rm N}\simeq 2.6$, indicating a suppression of $T_{\rm N}$ from 68\,K in the absence of these effects to the observed value of about 30\,K in BiMn$_2$PO$_6$.

\end{abstract}

%\keywords{frustration, manganites, NMR}

\maketitle

\section{Introduction}
\label{intro}
The antiferromagnetic (AFM) two-leg spin ladder is one of the most peculiar low-dimensional lattice topologies. Its properties are quite different from those of a simple spin chain, because the formation of rungs connecting the linear spin chains results in the dimerization and opens the spin gap or increases the size of the gap already existing for an isolated chain, thus protecting the system from a long-range magnetic order (LRO).\cite{johnston1996,batchelor2007} \mbox{Spin-$\frac12$} two-leg ladders have been extensively studied in the past. They show one-dimensional (1D) Luttinger liquid physics in high magnetic fields\cite{ward2013} and enjoy interesting connections to unconventional superconductivity that may emerge upon doping,\cite{dagotto1992} although experimental attempts to pursue this scenario in simple spin ladders have not been successful so far.\cite{hiroi1995} Sr$_{14}$Cu$_{24}$O$_{41}$ is the only two-leg spin ladder compound where superconductivity has been reported for hole doping at the Sr site under pressure.\cite{matsuda1996} Spin ladders with larger magnetic moments are relatively less studied. They feature weaker quantum fluctuations and, therefore, they are more likely to develop the LRO and conventional physics of classical antiferromagnets.\cite{kageyama2008} On the other hand, the larger energy of magnetic interactions in systems with high spin may be comparable to lattice energies and lead to intricate magnetostructural transitions, as in the spin-$\frac52$ ladder material BaMn$_2$O$_3$.\cite{valldor2011}

\begin{figure*}
\includegraphics{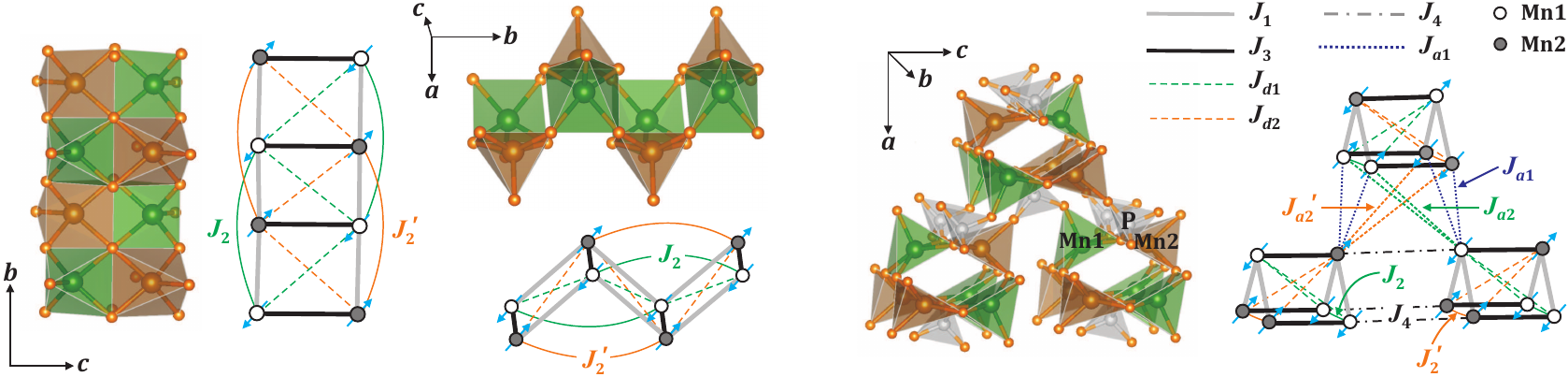}
\caption{\label{structure} (Color online)
Crystal structure of BiMn$_2$PO$_6$ and relevant magnetic interactions. Green, brown, and gray polyhedra show Mn1O$_5$, Mn2O$_5$, and PO$_4$, respectively. Bi atoms are not shown. Empty and filled circles denote the Mn1 and Mn2 positions, respectively. Left panel: different projections of the zigzag (buckled) ladder unit. Right panel: overall view of the structure. The antiferromagnetic classical ground state predicted by our electronic structure calculations is shown by arrows. The collinear ordering axis is chosen arbitrarily and may not reflect the actual ordering axis in the crystal. The exchange integrals ($J_{ij}$) in BiMn$_2$PO$_6$ and BiCu$_2$PO$_6$ are listed in Table~\ref{tab:exchanges} below: $J_1$ and $J_3$ are, respectively, leg and rung couplings of the zigzag ladder units. The crystal structures are visualized using the \texttt{VESTA} software.\cite{vesta}}
\end{figure*}

The family of Bi$M_2$PO$_6$ phosphates ($M$ is a transition-metal atom) hosts several interesting spin-ladder materials. Here, two $M$O$_5$ square pyramids containing $M$ atoms $M1$ and $M2$ in two different crystallographic positions share edges and form rungs of the ladder, as shown for BiMn$_2$PO$_6$ in Fig.~\ref{structure}.\cite{vesta}  These rungs connect to each other by corner-sharing of the $M$O$_5$ pyramids and build zigzag (buckled) two-leg ladders running along the $b$~direction. PO$_4$ tetrahedra connect the ladders and also form additional bridges within individual ladder units. The ensuing atomic arrangement is rather complex and may lead to multiple interactions beyond nearest-neighbor couplings along the leg ($J_1$) and along the rung ($J_3$) of the zigzag ladder.

Indeed, BiCu$_2$PO$_6$, which is the most actively studied member of the Bi$M_2$PO$_6$ family, reveals a highly non-trivial microscopic magnetic model.\cite{koteswararao2007,mentre2009,tsirlin2010} It does feature two-leg spin ladders consisting of two Cu chains connected by rung interactions, but the rung interactions are between the structural ladder units, so that rungs of the ladder are formed by the couplings $J_4$ (see Fig.~\ref{structure}), whereas the coupling $J_3$ turns out to be an interladder coupling.  Furthermore, the nearest-neighbor couplings $J_1$ are accompanied by next-nearest-neighbor couplings $J_2$ and $J_2^\prime$ that also run along the ladder (i.e., along $b$) and frustrate $J_1$, thus leading to a very intricate magnetic system.\cite{tsirlin2010}  So far, there is no clear consensus on whether BiCu$_2$PO$_6$ should be regarded as quasi-one-dimensional (1D) or quasi-two-dimensional (2D), i.e., whether the couplings $J_3$ (within the structural ladders, but between the spin ladders) are strong enough to build magnetic layers.\cite{tsirlin2010,plumb2013} BiCu$_2$PO$_6$ shows intriguing physical behavior,\cite{choi2013} especially in high magnetic fields, where multiple ordered phases emerge,\cite{kohama2012,casola2013} and upon doping with nonmagnetic (Zn$^{+2}$) or magnetic (Ni$^{+2}$) impurities.\cite{bobroff2009,casola2010}

Motivated by this interesting behavior, we studied the Mn$^{+2}$-based analog of BiCu$_2$PO$_6$. While the Cu$^{+2}$ compound features spin-$\frac12$ magnetic ions triggering strong quantum fluctuations, BiMn$_2$PO$_6$ (Ref.~\onlinecite{xun2002}) approaches the opposite limit of Mn$^{+2}$ spin-$\frac52$ cations that should be reasonably described by a classical Heisenberg model. The classical description might have simplified the microscopic analysis and given some clues about the puzzling magnetism of BiCu$_2$PO$_6$. Instead, we find that the replacement of Cu$^{+2}$ with Mn$^{+2}$ leads to a substantial change in the spin lattice, thus rendering BiCu$_2$PO$_6$ and BiMn$_2$PO$_6$ very different even on the level of individual interactions, let alone the ensuing magnetic behavior. In contrast to BiCu$_2$PO$_6$ with a gapped spin-liquid ground state and low-dimensional magnetic behavior, BiMn$_2$PO$_6$ is magnetically three-dimensional (3D), albeit with a pronounced 1D spatial anisotropy of exchange couplings. It develops long-range AFM order below about 30\,K and shows an additional magnetic transition around 10\,K\@. In the following, we report a comprehensive characterization of this material in terms of its structure, thermodynamic properties, microscopic magnetic model, magnetic ground state, and spin dynamics.

\section{Methods}

Polycrystalline samples of BiMn$_{2}$PO$_{6}$ and BiZn$_{2}$PO$_{6}$ were prepared by solid-state reaction techniques using Bi$_{2}$O$_{3}$ ($99.999$\%), MnO ($99.99$\%), ZnO ($99.99$\%), and NH$_{4}$H$_{2}$PO$_{4}$ ($99.9$\%) as starting materials, all from Sigma-Aldrich. The stoichiometric mixtures were heated at $800~^{\circ }$C in flowing Ar and in air with one intermediate grinding each for BiMn$_{2}$PO$_{6}$ and BiZn$_{2}$PO$_{6}$, respectively.

The resulting samples were single-phase as determined by x-ray diffraction (XRD, PANalytical powder diffractometer and CuK$_{\alpha}$ radiation, $\lambda_{\rm ave}=1.54182$~\AA) at room temperature. Le Bail profile fits to the XRD data were performed using the \texttt{Jana2006} software.\cite{bail}

Magnetic susceptibility $\chi \equiv M/H$ data were measured versus temperature $T$ and applied magnetic field $H$ using a SQUID magnetometer [Quantum Design, Magnetic Properties Measurement System (MPMS)]. Heat capacity $C_{\rm p}$ data were collected with a Quantum Design Physical Properties Measurement System (PPMS) on pressed pellets using the relaxation technique.

The nuclear magnetic resonance (NMR) measurements were carried out using pulsed NMR techniques on $^{31}$P nuclei with spin $I=\frac12$ and gyromagnetic ratio $\bar{\gamma}_N =\gamma_N/2\pi = 17.237$~MHz/Tesla, over the $T$ range \mbox{$4$~K $\leq T \leq300$~K}\@.  The NMR measurements were done at two radio frequencies of 77.5~MHz and 49.15~MHz. Spectra were obtained either by Fourier transform (FT) of the NMR echo signal or by sweeping the field at fixed frequency.  The NMR shift $K(T)=\left[ H_{\text{ref}}-H\left( T\right)\right]/H(T)$ was determined by measuring the resonance field $H\left(T\right)$ of the sample with respect to a standard H$_{3}$PO$_{4}$ solution (resonance field $H_{\text{ref}}$). The $^{31}$P nuclear spin-lattice relaxation rate
$(1/T_{1})$ was measured after applying a comb of saturation pulses.

Individual magnetic couplings in BiMn$_2$PO$_6$ were evaluated from density-functional theory (DFT) band-structure calculations performed in the \texttt{FPLO} code\cite{fplo} within the generalized gradient approximation (GGA)\cite{pbe96} augmented by a mean-field correction for correlation effects in the Mn $3d$ shell (GGA+$U$). We used the on-site Coulomb repulsion parameter $U_d=5.5$\,eV and the on-site Hund's coupling $J_d=1$\,eV that yield exchange integrals in quantitative agreement with the experimental data. While no conclusive information on the values of $U_d$ and $J_d$ appropriate for Mn$^{+2}$ is available in the literature, we note that our choice of $U_d=5.5$\,eV is compatible with earlier computational studies, where $U_d=4$--6\,eV has been used.\cite{kasinathan2006,johnson2013} The variation of $U_d$ in the $4$--6\,eV range leads to marginal changes in the computed exchange integrals $J_{ij}$, with less than 10\% variation in the absolute values. Each $J_{ij}$ was evaluated from total energies of four collinear magnetic configurations, as described in Ref.~\onlinecite{xiang2011}.

The magnetic susceptibility and ground state of the DFT-based magnetic model were evaluated by the classical Monte-Carlo \texttt{spinmc} algorithm of the \texttt{ALPS} simulation package.\cite{alps} Additionally, we used the quantum Monte-Carlo \texttt{loop} algorithm\cite{todo2001} for the non-frustrated reference model considered in Sec.~\ref{sec:microscopic}. Simulations were performed for finite lattices with periodic boundary conditions and up to 4096 sites. Convergence with respect to finite-size effects was carefully checked. 

\section{Results}

\subsection{Crystallography}

\begin{figure}
\includegraphics[scale=0.97]{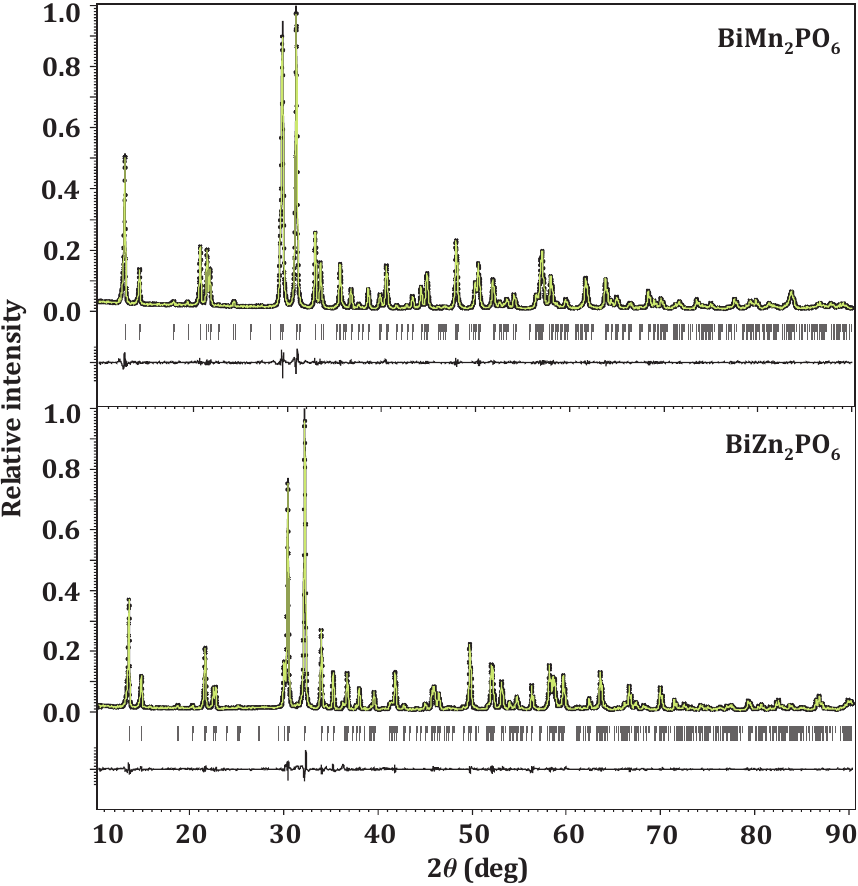}
\caption{\label{XRD} (Color online) Le Bail fits of x-ray powder diffraction patterns of BiMn$_{2}$PO$_{6}$ (upper panel) and BiZn$_{2}$PO$_{6}$ (lower panel). In each graph, the vertical tick marks show the Bragg  reflection positions, which are duplicated according to the mixed CuK$_{\alpha1/\alpha2}$ radiation, and the bottom line is the difference curve.}
\end{figure}

\begin{table}[ptb]
\caption{Crystallographic data for BiMn$_2$PO$_6$ at room temperature (orthorhombic structure, space group $Pnma$).\cite{xun2002} Our fitted lattice parameters are $a=12.0383(2)$\,\r A, $b=5.3656(1)$\,\r A, and $c=8.1207(1)$\,\r A compared to the reported values $a=12.0425(4)$\,\r A, $b=5.3704(1)$\,\r A, and $c=8.1288(2)$\,\r A.\cite{xun2002} Our goodness-of-fit is obtained to be $R_{\rm p} = 4.9\%$. Listed are the Wyckoff symbols and relative atomic coordinates $x/a$, $y/b$, and $z/c$ of each atom.\cite{xun2002}}
\label{Cry_parameters1}
\begin{ruledtabular}
\begin{tabular}{cccccc}
Atom & Wyckoff position & $x/a$ & $y/b$ & $z/c$ \\\hline
Bi  & $4c$ & $0.0950(2)$ & $1/4$ & $0.0120(5)$ \\
Mn1 & $4c$ & $0.1032(6)$ & $3/4$ & $0.6924(6)$ \\
Mn2 & $4c$ & $0.0991(7)$ & $3/4$ & $0.2952(7)$ \\
P   & $4c$ & $0.1970(3)$ & $1/4$ & $0.4744(7)$ \\
O1  & $8d$ & $-0.0033(3)$ & $0.0050(7)$ & $0.1634(2)$ \\
O2  & $8d$ & $0.1249(2)$ & $0.4859(4)$ & $0.4922(5)$ \\
O3  & $4c$ & $0.2895(4)$ & $1/4$ & $0.5983(6)$ \\
O4  & $4c$ & $0.2414(3)$ & $1/4$ & $0.2965(5)$ \\
\end{tabular}
\end{ruledtabular}
\end{table}

\begin{table}[ptb]
\caption{Crystallographic data for BiZn$_2$PO$_6$ at room temperature (primitive orthorhombic structure, space group $Pnma$).\cite{ketatni2000} Our fitted lattice parameters are $a=11.8941(2)$\,\r A, $b=5.2753(1)$\,\r A, and $c=7.8150(1)$\,\r A compared to the reported values $a=11.8941(3)$\,\r A, $b=5.2754(2)$\,\r A, and $c=7.8161(2)$\,\r A.\cite{ketatni2000} Our goodness-of-fit is obtained to be $R_{\rm p} = 6.1\%$. Listed are the Wyckoff symbols and relative atomic coordinates $x/a$, $y/b$, and $z/c$ of each atom.\cite{ketatni2000}}
\label{Cry_parameters2}
\begin{ruledtabular}
\begin{tabular}{cccccc}
Atom & Wyckoff position & $x/a$ & $y/b$ & $z/c$ \\\hline
Bi  & $4c$ & $0.0990(2)$ & $1/4$ & $0.0119(3)$ \\
Zn1 & $4c$ & $0.1028(7)$ & $3/4$ & $0.6915(6)$ \\
Zn2 & $4c$ & $0.0930(7)$ & $3/4$ & $0.3011(6)$ \\
P   & $4c$ & $0.1945(8)$ & $1/4$ & $0.481(2)$ \\
O1  & $8d$ & $-0.010(2)$ & $-0.006(4)$ & $0.191(2)$ \\
O2  & $8d$ & $0.123(1)$ & $0.497(2)$ & $0.489(3)$ \\
O3  & $4c$ & $0.285(1)$ & $1/4$ & $0.604(2)$ \\
O4  & $4c$ & $0.245(2)$ & $1/4$ & $0.315(3)$ \\
\end{tabular}
\end{ruledtabular}
\end{table}

BiMn$_{2}$PO$_{6}$ and its nonmagnetic sibling BiZn$_{2}$PO$_{6}$ crystallize in the primitive orthorhombic space group $Pnma$ (No. 62) containing $Z = 4$ formula units per unit cell.  The crystal structures were solved in Refs.~\onlinecite{xun2002} and \onlinecite{ketatni2000} using neutron and x-ray powder diffraction, respectively.  The atomic positions determined by these authors for the respective compounds are given in Tables~\ref{Cry_parameters1} and~\ref{Cry_parameters2}, and the lattice parameters in the respective figure captions.  These compounds are isostructural to BiCu$_2$PO$_6$.\cite{abraham1994}

We carried out powder x-ray diffraction measurements of our polycrystalline samples of BiMn$_{2}$PO$_{6}$ and BiZn$_{2}$PO$_{6}$ and the results are shown in Fig.~\ref{XRD}. Le Bail fits of the patterns based on space group $Pnma$ were done to determine the lattice parameters.  Good fits were obtained as shown in Fig.~\ref{XRD}, and the respective lattice parameters are listed in the captions of Tables~\ref{Cry_parameters1} and~\ref{Cry_parameters2}. Excellent agreement of our lattice parameters with those previously determined for the two compounds is seen in the respective table captions.

Details of this crystal structure (Fig.~\ref{structure}) have been discussed in Sec.~\ref{intro}. The most notable difference between the Mn$^{+2}$, Cu$^{+2}$, and Zn$^{+2}$ compounds lies in the geometry of the $M$O$_5$ polyhedra. The CuO$_5$ square pyramids feature a 4+1 coordination, with 4 shorter in-plane distances of 1.9--2.0\,\r A forming a CuO$_4$ plaquette and the fifth apical distance of 2.20--2.35\,\r A.\cite{abraham1994} This 4+1 type of coordination is clearly reminiscent of the Jahn-Teller distortion of Cu$^{+2}$. Neither Mn$^{+2}$ nor Zn$^{+2}$ show this type of distortion. Their $M$O$_5$ polyhedra are more regular, with all five $M$--O distances lying in the range of 2.05--2.17\,\r A for Mn$^{+2}$ (Ref.~\onlinecite{xun2002}) and 1.97--2.12\,\r A for the smaller Zn$^{+2}$ cation.\cite{ketatni2000}

\subsection{\textbf{Magnetization and Magnetic Susceptibility}}
\label{sec:magnetization}

\begin{figure}
\includegraphics [width=8.6cm] {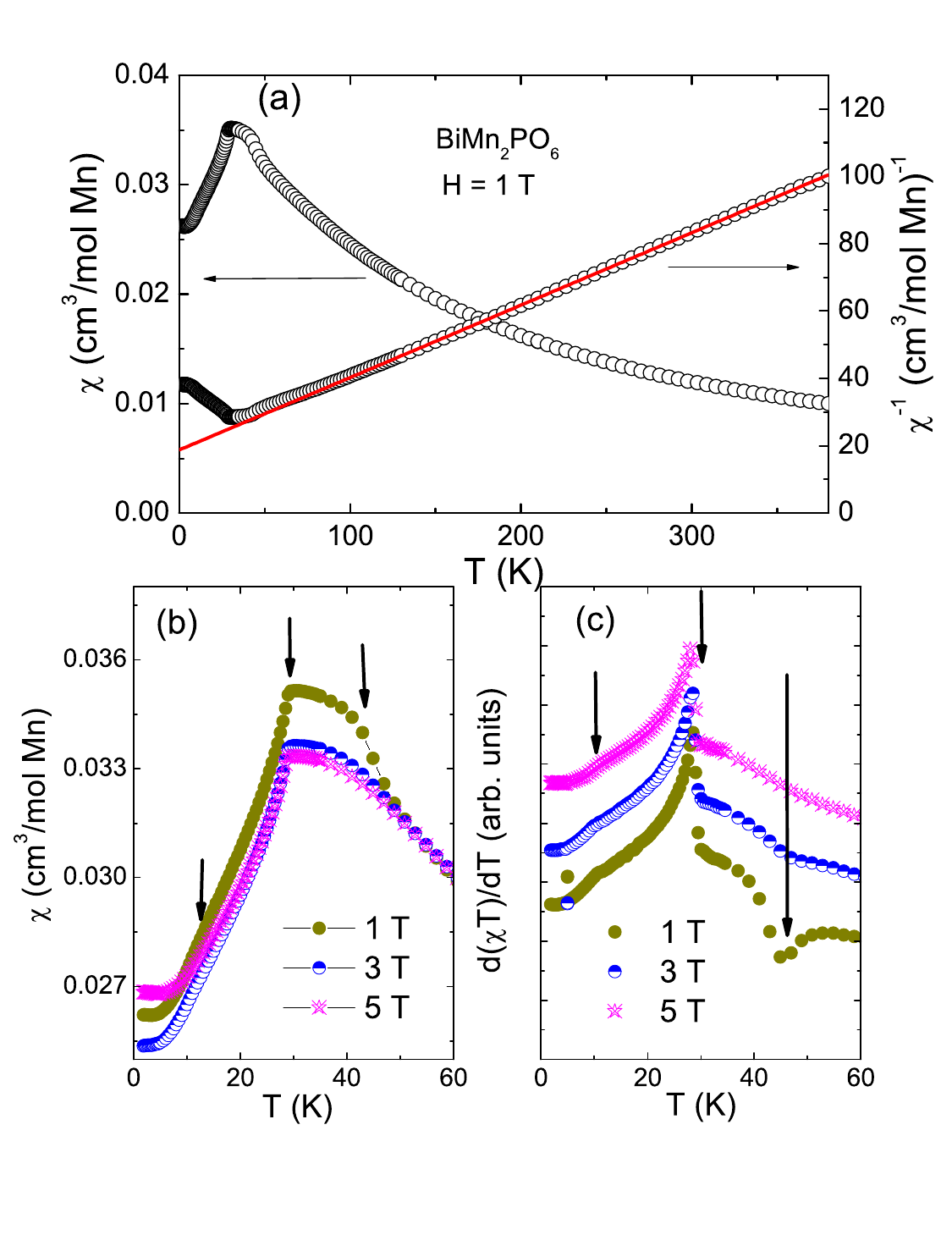}
\caption{\label{chi} (Color online) (a) The magnetic susceptibility $\chi$ and inverse magnetic susceptibility $\chi^{-1}$ of BiMn$_{2}$PO$_{6}$ versus $T$ measured at an applied magnetic field $H=1$~T are plotted along left and right $y$-axes, respectively. The straight red line is a CW fit of $\chi^{-1}(T)$ from 150 to 350~K\@. (b) $\chi$ versus $T$ measured at three different applied fields in the
low-$T$ regime. (c) $\partial(\chi T)/\partial T$ versus $T$ for three different applied fields in the
low-$T$ regime. The 3\,T and 5\,T data are offset vertically by 0.01 and 0.02~cm$^3$K/mol~Mn, respectively. In panels (b) and (c), the downward-pointing arrows indicate magnetic transitions. The one at 43~K is extrinsic, likely originating from a Mn$_3$O$_4$ impurity phase.}
\end{figure}

The magnetic susceptibility $\chi(T)$ data for BiMn$_{2}$PO$_{6}$
measured at $H$ = 1\,T are presented in Fig.~\ref{chi}. At high
temperatures $T>150$~K, $\chi(T)$ follows a Curie-Weiss law.  With decrease in
temperature, a sudden jump at 43\,K, a peak at 30\,K, and then a change in
slope at 10\,K were observed in $\chi(T)$ suggesting that there
are three possible magnetic transitions at low temperatures as noted by the vertical arrows for $H$ = 1~T in Fig.~3(b). No broad maximum associated with dynamic short-range AFM ordering was observed down to low temperatures.

To fit the uniform magnetic susceptibility data at high temperatures, we used the expression
\begin{equation}
\chi=\chi_{0}+\frac{C}{T-\theta_{\rm CW}},
\label{Curie}
\end{equation}
where $\chi_{0}$ is the temperature-independent contribution that accounts for core diamagnetism and Van Vleck (VV) paramagnetism.  The second term is the Curie-Weiss (CW) law with Curie constant
$C$ and Weiss temperature $\theta_{\rm CW}$.  The data above 150\,K were fitted with the parameters
$\chi_{0} = 4(3)\times 10^{-4}$ cm$^{3}$/mol~Mn, $C = 4.4(3)$~cm$^{3}$~K/mol~Mn, and $\theta_{\rm CW} = -78(7)$~K\@. The error bars were determined by varying the fitted temperature range. This value of $C$ is in good agreement with the value $C=4.377$~cm$^{3}$~K/mol~Mn for the high-spin state ($S=\frac52$) of Mn$^{+2}$ with $g$-factor $g=2.00$, as expected for Mn$^{+2}$.\cite{smith1968} Adding the core diamagnetic susceptibility for the individual ions ($\chi_{\text{Bi}^{+3}}=-25\times 10^{-6}$\,cm$^3$/mol, $\chi_{\text{Mn}^{+2}}=-14\times 10^{-6}$\,cm$^3$/mol, $\chi_{\text{P}^{+5}}=-1\times 10^{-6}$\,cm$^3$/mol, and $\chi_{\text{O}^{-2}}=-12\times 10^{-6}$\,cm$^3$/mol),\cite{Selwood1956} the total $\chi_{\rm{core}}$ was calculated to be $-1.26\times 10^{-4}$\,cm$^3$/mol. The Van Vleck paramagnetic susceptibility for BiMn$_{2}$PO$_{6}$ estimated by subtracting $\chi_{\rm{core}}$ from $\chi_0$ is $\chi_{\rm{VV}}\simeq 4.6\times 10^{-4}$\,cm$^3$/mol~Mn.
The large negative value of $\theta_{\rm CW}$ shows that the dominant interactions between the Mn spins are AFM\@. Below 150\,K, the $1/\chi$ data in Fig.~\ref{chi}(a) begin to deviate from the CW fit, which suggests the onset of AFM correlations beyond those described by the Curie-Weiss law.

In order to further confirm the sequence of magnetic transitions at low temperatures,
$\chi(T)$ was also measured at different applied fields. As seen in
Fig.~\ref{chi}(b), the sudden jump at 43\,K observed at $H=1$\,T is completely
suppressed at $H=3$\,T. On the other hand, the peak at 30\,K and the bump at 10\,K are not affected at all by
external fields up to 5\,T. In a simple antiferromagnet, the magnetic specific heat ($C_{\rm mag}$) is related to the parallel static uniform susceptibility $\chi$ by Fisher's relation\cite{fisher1962} for AFMs given by
\begin{equation}
C_{\rm mag} \simeq A\,\frac{\partial(\chi T)}{\partial T},
\label{fisher}
\end{equation}
where the proportionality factor $A$ is expected to be a slowly varying function of $T$ near $T_{\rm N}$. This relation has been verified experimentally for some bulk materials.\cite{bragg1973} For clarity, we have plotted the $T$-derivative
of $\chi T$ as a function of $T$ in Fig.~\ref{chi}(c) measured at three different applied fields. Figure~\ref{chi}(c) confirms that the transitions at 30\,K and 10\,K remain unchanged for $H$ up to 5~T. The feature at 43\,K is likely due to the
presence of Mn$_{3}$O$_{4}$ impurity phase, which orders ferrimagnetically at 42~K.\cite{dwight1960,chardon1986} While we do not see this impurity in x-ray powder diffraction data (Fig.~\ref{XRD}), even a trace amount of Mn$_3$O$_4$ (below 1\,\%) may be sufficient to produce a visible magnetic anomaly around 43\,K.

\begin{figure}
\includegraphics[width=3.3in]{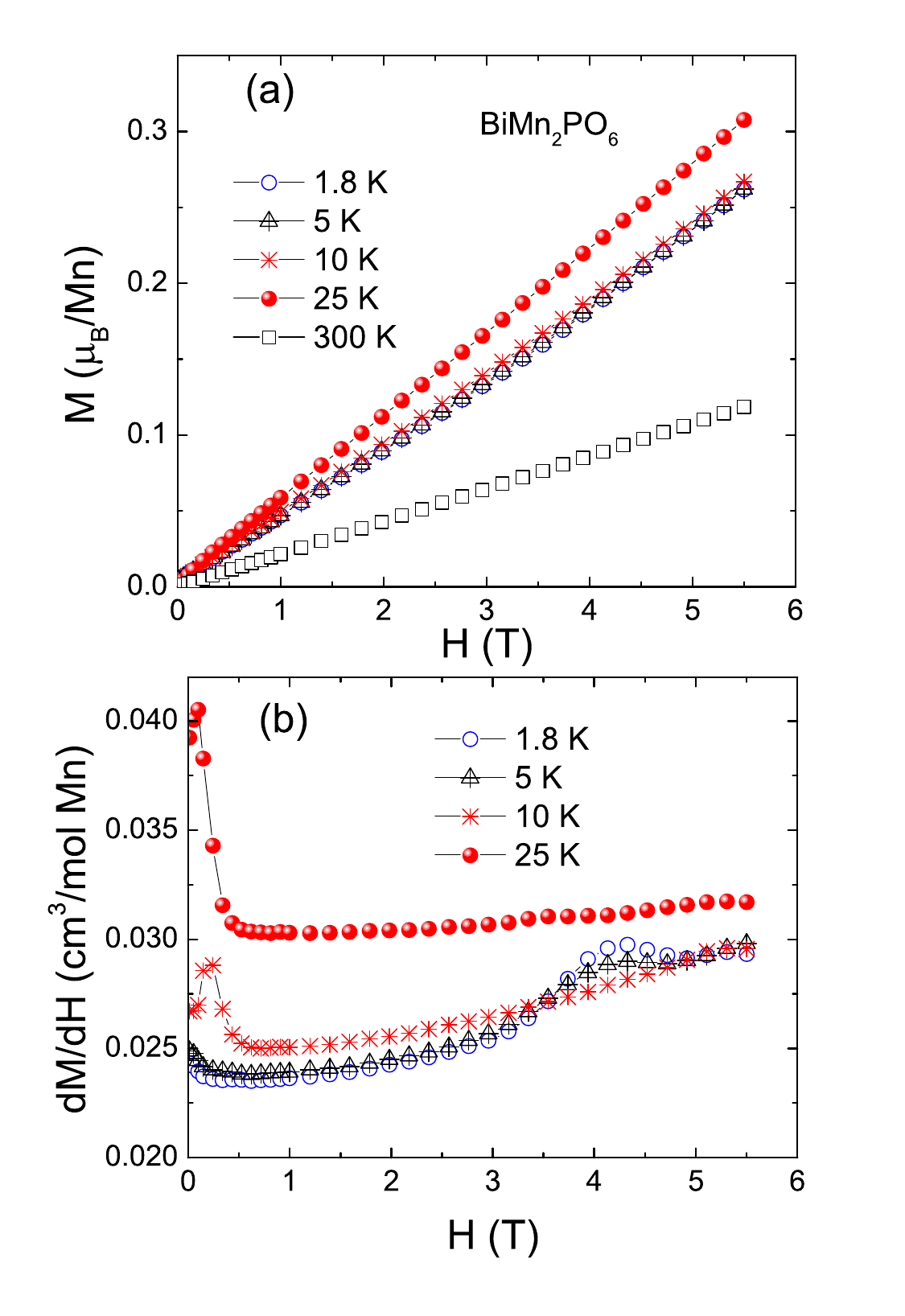}
\caption{\label{MH} (Color online) (a) Magnetization ($M$) as a function of
applied field ($H$) measured at different temperatures. (b)~Field derivative
of magnetization ($dM/dH$) at 1.8\,K, 5\,K, 10\,K, and 25\,K as a function of $H$ to
highlight the field-induced transition at $\approx 4.5$~T, which appears to be
a property of the low-$T$ ($<10$~K) magnetic phase. The peak in $dM/dH$ at $\sim0.2$~T
in~(b) arises from a Mn$_3$O$_4$ impurity phase (see text).}
\end{figure}

The susceptibility of BiMn$_2$PO$_6$ in Fig.~\ref{chi}(a) is typical for a 3D antiferromagnet. In particular, we do not observe a broad maximum above $T_{\rm N}$ that would be expected for a quasi-1D system, as in BaMn$_2$O$_3$.\cite{valldor2011} On the other hand, as shown below the individual magnetic couplings are somewhat anisotropic, and this spatial anisotropy together with frustration effects have a pronounced influence on $T_{\rm N}$. Details of the microscopic magnetic model are discussed in Secs.~\ref{sec:dft} and~\ref{sec:microscopic}.

$M(H)$ isotherms were measured at different temperatures, shown in Fig.~\ref{MH}(a), to
check for field-induced effects and for the presence of the probable ferrimagnetic Mn$_3$O$_4$ impurity
phase in the sample. Above 50\,K, $M$ is proportional to $H$ over the whole
field range. At 25\,K, a nonlinearity was observed in the $M(H)$ curve below
about 0.3\,T suggesting a small ferrimagnetic Mn$_3$O$_4$ impurity contribution
in the magnetization. In order to quantitatively estimate the Mn$_3$O$_4$
impurity concentration, we fitted the $M(H)$ isotherm at 25~K in the field range
1~T to 5.5~T by the linear relation $M(H)=M_{\rm s}+\chi H$, where $M_{\rm s}$
is the saturation magnetization of the Mn$_3$O$_4$ ferrimagnetic impurity
and $\chi$ is the intrinsic magnetic susceptibility of the sample. The
obtained value of $M_{\rm s} \simeq 0.00492~\mu_{\rm B}$/f.u. corresponds to about
$0.26$~mol\,\% Mn$_3$O$_4$ impurity [$M_{\rm s}=1.87~\mu_{\rm B}$/f.u. for Mn$_3$O$_4$
at $T=0$~K].\cite{dwight1960} This small amount is not observable from our x-ray
diffraction measurements. At 1.8\,K, in the maximum field of 5.5\,T, $M \simeq
0.26~\mu_{\rm B}$/Mn is reached, which corresponds to only 5\% of the fully
polarized magnetization of 5~$\mu_{\rm B}$/Mn. This agrees with a dominant
antiferromagnetic exchange coupling in BiMn$_{2}$PO$_{6}$.

To further elucidate the dependence of $M$ on $H$, shown in Fig.~\ref{MH}(b) is a plot
of $dM/dH$ versus~$H$ at 1.8, 5, 10, and~25~K\@. A sharp peak is observed at a
low field of $\sim 0.1$~T at 25~K that we attribute to the saturation of the
${\rm Mn_3O_4}$ impurity phase that has a Curie temperature of 43~K\@.  At 25~K,
the integral of $dM/dH$ versus~$H$ from $H = 0$ to~0.5~T is $\sim39$~G~cm$^3$/mol,
which is comparable with the value of $M_{\rm s}$ that we obtained above. We
believe that this peak becomes smaller with decreasing~$T$ and disappears below
10~K because the thermal energy cannot easily overcome the anisotropy energy of
the ferrimagnetic domain walls in ${\rm Mn_3O_4}$ at low fields with decreasing~$T$\@.
Therefore the saturation of the ferrimagnetic component of ${\rm Mn_3O_4}$ occurs
over a wide field range, resulting in a strong decrease in the height of the
low-field peak in $dM/dH$ with decreasing~$T$ below 25~K\@.

In addition to the above extrinsic low-field peak in $dM/dH$ versus~$H$ arising
from the ${\rm Mn_3O_4}$ impurity phase, we also observe an intrinsic high-field
metamagnetic transition at $H\approx 4.5$~T [see Fig.~\ref{MH}(b)]. This transition is not seen at
10~K and therefore likely pertains to the low-$T$ magnetic phase below 10~K only. We refrain
here from speculating on the nature of this metamagnetic transition because the
magnetic structure below 10~K is not yet known.

\subsection{Heat Capacity}
\label{sec:heat}

\begin{figure}[t]
\includegraphics[width=3.3in]{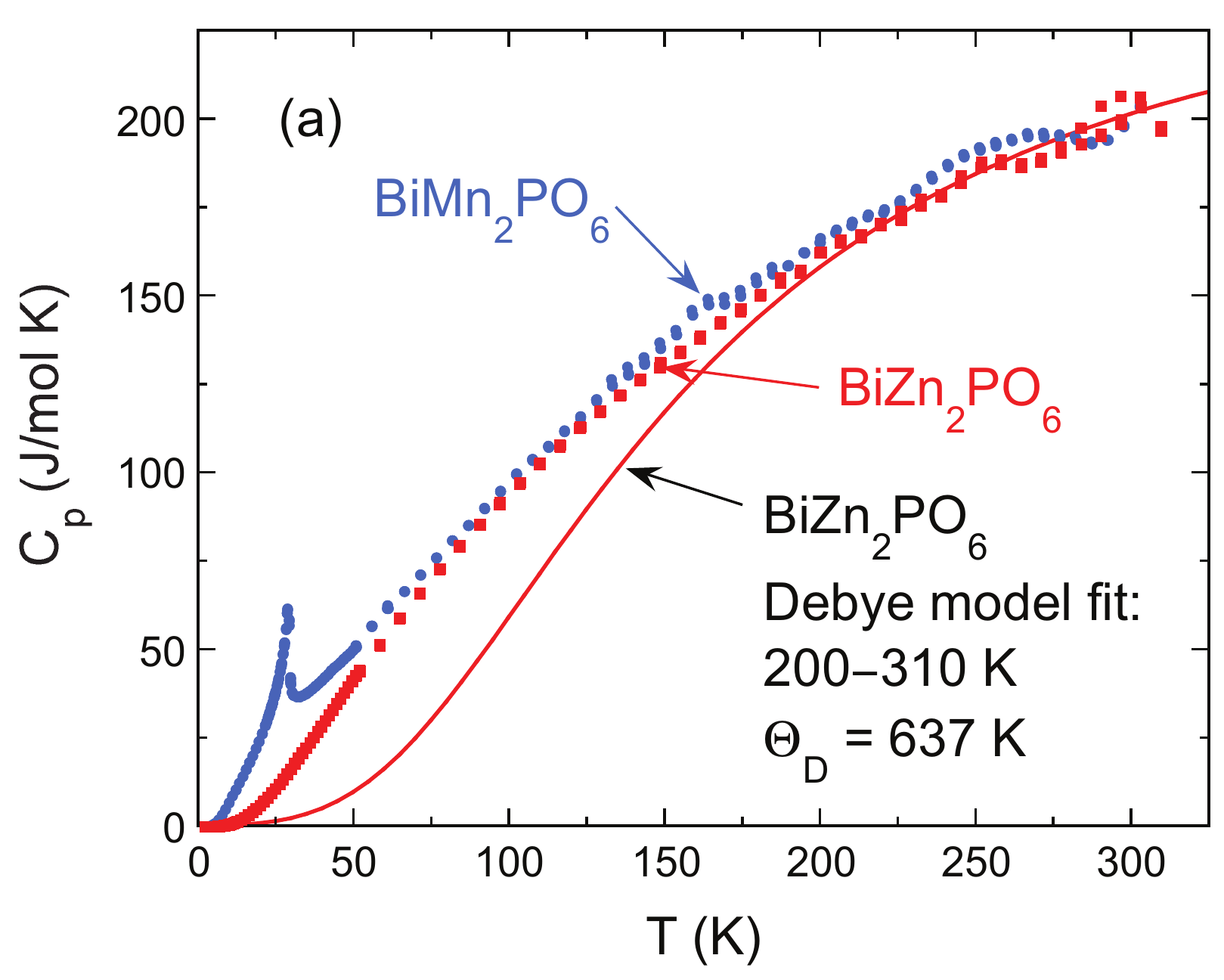}
\includegraphics[width=3.3in]{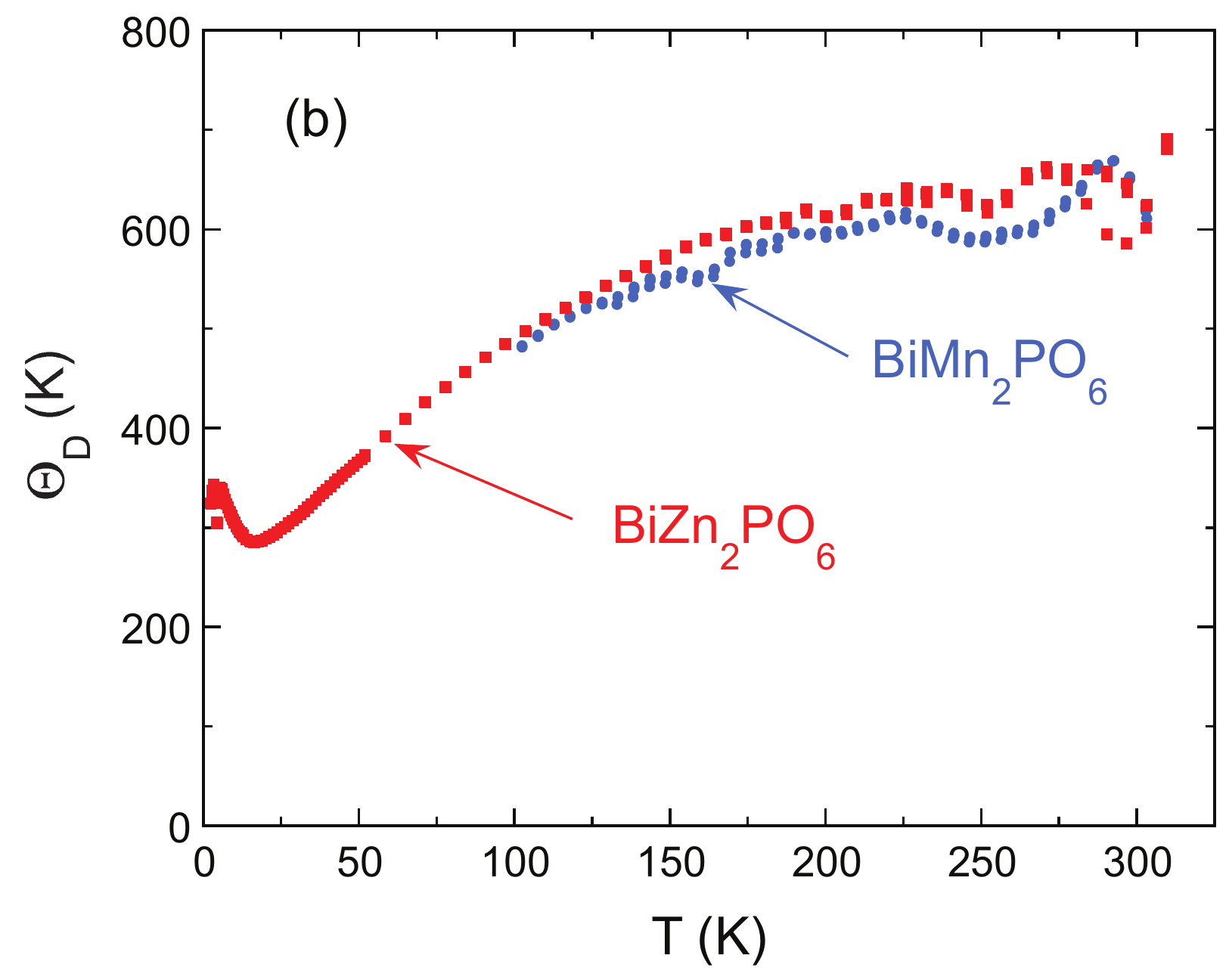}
\caption{(Color online)  (a) Overview of the heat capacity at constant pressure $C_{\rm p}$ versus temperature~$T$ of ${\rm BiMn_2PO_6}$ (filled blue circles) and of the nonmagnetic analogue ${\rm BiZn_2PO_6}$ (filled red squares) from 2 to 310~K\@. Also shown is a fit of the Debye heat capacity model in Eq.~(\ref{Eq:CVDebye}) to the data for nonmagnetic ${\rm BiZn_2PO_6}$ between 200 and 310~K, which yields the Debye temperature $\Theta_{\rm D} = 637$~K\@. The small dips and bumps in the data for $T\sim 275$~K are believed to be artifacts. (b) Debye temperature $\Theta_{\rm D}$ versus $T$ computed using Eq.~(\ref{Eq:CVDebye}) from the individual data points for the two compounds in~(a). Only data above 100~K are shown for ${\rm BiMn_2PO_6}$ because of the additional magnetic contribution below this $T$\@. The lattices of both compounds show a drastic softening on cooling below $\sim200$~K.}
\label{Fig:BiMnZn2PO6_Cp}
\end{figure}

An overview of the $C_{\rm p}(T)$ data for ${\rm BiMn_2PO_6}$ and the nonmagnetic analogue ${\rm BiZn_2PO_6}$ from 2 to 310~K is shown in Fig.~\ref{Fig:BiMnZn2PO6_Cp}(a).  A sharp $\lambda$-type anomaly is seen for ${\rm BiMn_2PO_6}$ at $T\approx 29$~K associated with the above long-range AFM order, discussed in more detail below.  The Debye model for the lattice heat capacity at constant volume $C_{\rm V}$ arising from acoustic phonons is given by\cite{kittel1966}
\be
\frac{C_{\rm V}}{nR} = 9\left(\frac{T}{\Theta_{\rm D}}\right)^3\int_0^{\Theta_{\rm D}/T}\frac{x^4e^x}{(e^x-1)^2}\,dx,
\label{Eq:CVDebye}
\ee
where $R$ is the molar gas constant, $n$ is the number of atoms per formula unit, and $\Theta_{\rm D}$ is the Debye temperature. This prediction was recently accurately fitted by an analytic Pad\'e approximant which greatly simplifies fitting experimental data by Eq.~(\ref{Eq:CVDebye}).\cite{Goetsch2012}

We fitted the $C_{\rm p}(T)$ data for nonmagnetic ${\rm BiZn_2PO_6}$ over the full temperature range by Eq.~(\ref{Eq:CVDebye}) using the  Pad\'e approximant formulation, but the fit was poor.  A better fit was obtained to the data just from 200 to 310~K, as shown by the red curve in Fig.~\ref{Fig:BiMnZn2PO6_Cp}(a), where the fitted Debye temperature is $\Theta_{\rm D} = 637$~K\@.  The large  $\Theta_{\rm D}$ is typical of oxides due to the low mass of the O atoms and the strong interatomic bonding involving those atoms.  The fit strongly deviates from the data on cooling below $\sim200$~K, which we attribute to anomalous and strong {\it softening} of the lattice on cooling.

To quantify this lattice softening, the $\Theta_{\rm D}$ versus $T$ was calculated for each data point for the two compounds using the Pad\'e approximant formulation of Eq.~(\ref{Eq:CVDebye}) and the results are shown in Fig.~\ref{Fig:BiMnZn2PO6_Cp}(b), where only the data above 100~K are plotted for ${\rm BiMn_2PO_6}$ because as shown below the magnetic contribution to the heat capacity starts to become significant below this temperature.  As seen in Fig.~\ref{Fig:BiMnZn2PO6_Cp}(b), $\Theta_{\rm D}$ decreases by a factor of about two on cooling from 300~K to 2~K\@.  This is an extremely large change for solids, where the temperature variations below 300~K are typically $\pm20$\% due to differences between the actual phonon densities of states and that assumed in the Debye theory.\cite{Gopal1966}

\begin{figure}
\includegraphics[width=3.3in]{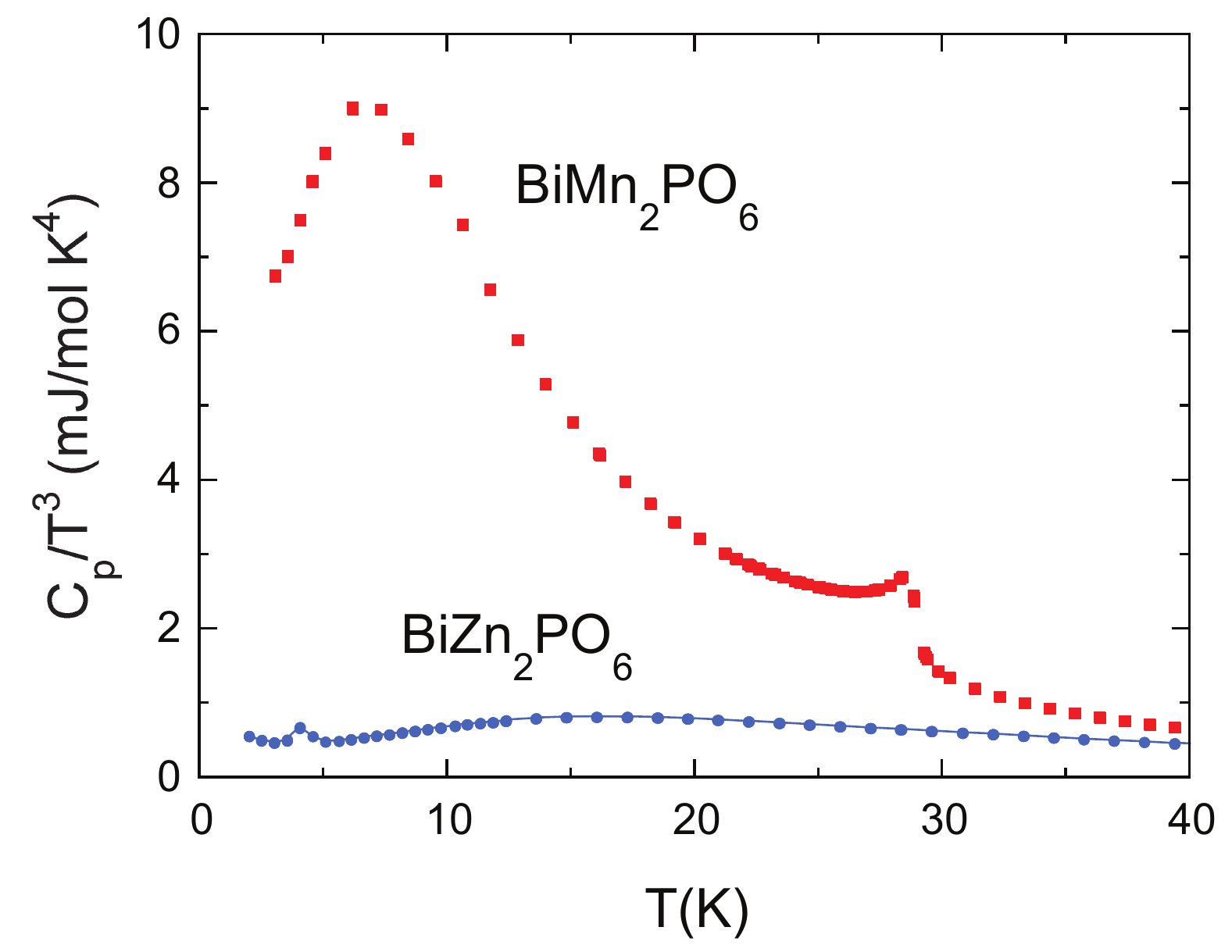}
\caption{(Color online)  Heat capacity $C_{\rm p}/T^3$ versus temperature~$T$ for ${\rm BiMn_2PO_6}$ and ${\rm BiZn_2PO_6}$ below 40~K\@.  For nonmagnetic ${\rm BiZn_2PO_6}$, the value at low~$T$ is the coefficient $\beta$ in the Debye $T^3$ law~(\ref{Eq:DebyeT3}) for the lattice heat capacity.  The large enhancement of $C_{\rm p}/T^3$ at low~$T$ for ${\rm BiMn_2PO_6}$ is likely of magnetic origin. }
\label{Fig:BiMnZn2PO6CpOnT3}
\end{figure}

Expanded plots of $C_{\rm p}/T^3$ versus $T$ for ${\rm BiMn_2PO_6}$ and ${\rm BiZn_2PO_6}$ below 40~K are shown in Fig.~\ref{Fig:BiMnZn2PO6CpOnT3}.  The low-$T$ limit of the Debye theory prediction in Eq.~(\ref{Eq:CVDebye}) is the so-called Debye~$T^3$ law, given by\cite{kittel1966}
\be
\frac{C_{\rm V}(T\to0)}{nR} = \frac{12\pi^4}{5}\left(\frac{T}{\Theta_{\rm D}}\right)^3 \equiv \beta T^3.
\label{Eq:DebyeT3}
\ee
For nonmagnetic ${\rm BiZn_2PO_6}$, Fig.~\ref{Fig:BiMnZn2PO6CpOnT3} gives $\beta\approx 0.45$~mJ/mol\,K$^4$ at low temperatures.  Then using $n=10$ atoms per formula unit, Eq.~(\ref{Eq:DebyeT3}) gives the Debye temperature as $\Theta_{\rm D}\approx 350$~K, consistent with the data for $T\to0$ in the point-by-point plot of $\Theta_{\rm D}(T)$ in Fig.~\ref{Fig:BiMnZn2PO6_Cp}(b).  From Fig.~\ref{Fig:BiMnZn2PO6CpOnT3}, one sees a large enhancement of $C_{\rm p}(T)$ for ${\rm BiMn_2PO_6}$ above that of ${\rm BiZn_2PO_6}$ at low temperatures.  This enhancement presumably originates from the magnetic degrees of freeedom (spin waves) in the AFM ordered state below $T_{\rm N}\approx 30$~K, which in turn indicates that any energy gap  in the spin-wave spectrum induced by magnetic anisotropy is negligible for $T \gtrsim 3$~K\@.  This topic is discussed in more detail in Sec.~\ref{Sec:SpinWaveCp}.

\begin{figure}
\includegraphics[width=3.3in]{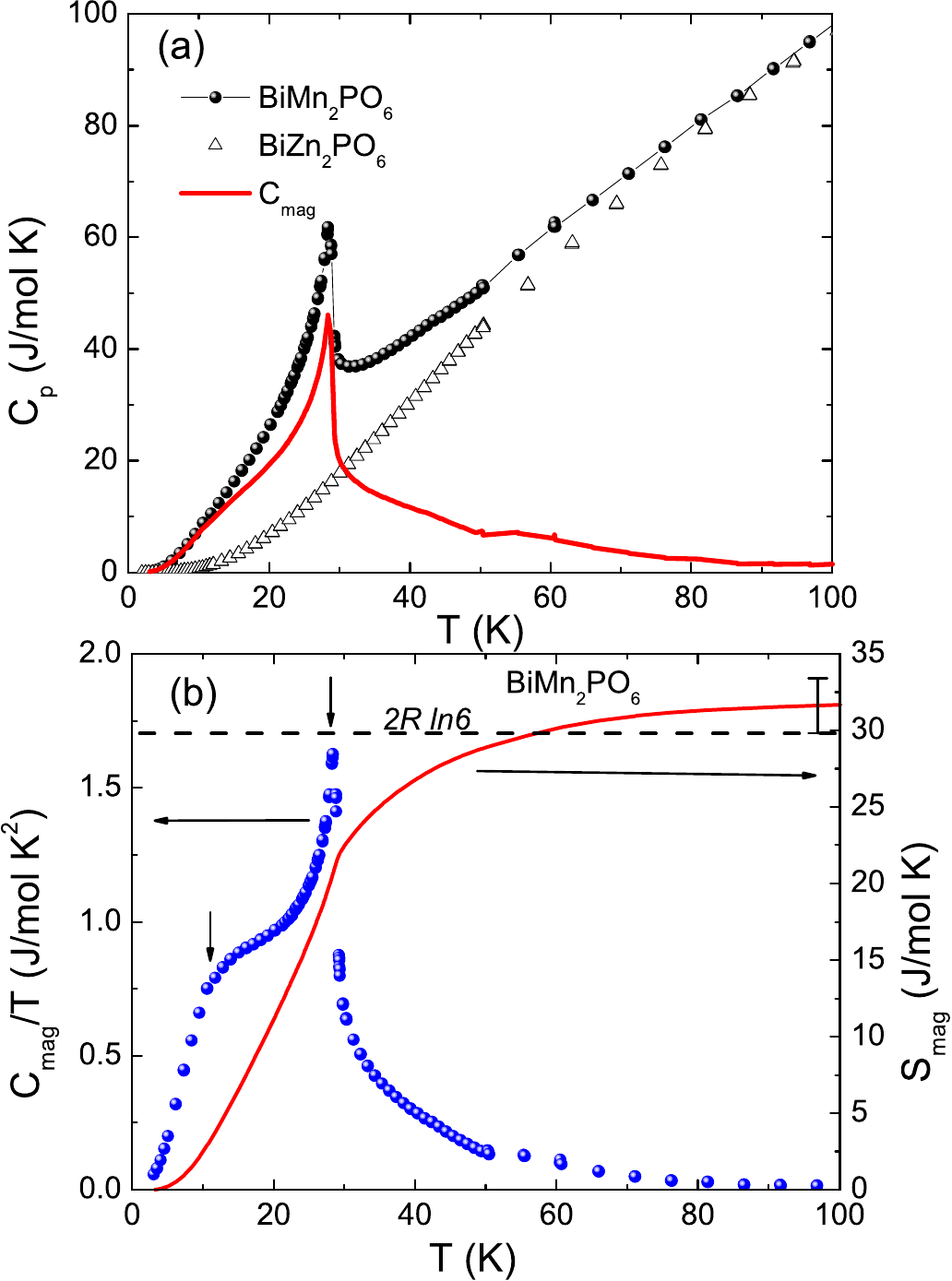}
\caption{(Color online) (a) Heat capacity $C_{\rm p}$ versus temperature $T$ for BiMn$_{2}$PO$_{6}$ and the nonmagnetic reference  compound BiZn$_{2}$PO$_{6}$. The red solid curve is the derived magnetic heat capacity
$C_{\rm mag}(T)$. (b) $C_{\rm mag}(T)/T$ and the magnetic entropy $S_{\rm mag}$ as a function of $T$ along the left and right $y$-axes, respectively. The dashed horizontal line is the value $S_{\rm mag} = 2R\ln 6$ expected per mole of f.u.\ for Mn$^{+2}$ ($S=\frac52$) spins.  The downward arrows indicate the two transition points. However, the broad peak in $C_{\rm mag}/T$ at $T\approx10$~K is not associated with a magnetic transition (see text).}
\label{cp}
\end{figure}

The $C_{\rm p}(T)$ data up to 100~K are shown in Fig.~\ref{cp}(a) for both ${\rm BiMn_2PO_6}$ and ${\rm BiZn_2PO_6}$ where the temperature scale of the latter data was corrected for the difference in formula weights of the two compounds.  With decreasing~$T$, the magnitude of the negative slope of $C_{\rm p}(T)$ for ${\rm BiMn_2PO_6}$ increases before a sharp $\lambda$-type anomaly occurs with a peak at the long-range AFM ordering temperature $T_{\rm N}\approx 29$~K\@.  In order to obtain a quantitative estimate of the magnetic contribution $C_{\rm mag}(T)$ to $C_{\rm p}(T)$, the mass-corrected $C_{\rm p}(T)$ of ${\rm BiZn_2PO_6}$ was subtracted from the measured data for ${\rm BiMn_2PO_6}$.  The resulting $C_{\rm mag}(T)$ is shown as the red curve in Fig.~\ref{cp}(a).  There is no broad peak in $C_{\rm mag}$ at $T>T_{\rm N}$, which suggests that the Mn--Mn exchange interaction connectivity in ${\rm BiMn_2PO_6}$ is essentially three-dimensional.  There is also no trace of a transition at 43~K, further supporting the extrinsic nature of the feature observed above in $\chi$ at about this~$T$\@.

The Mn ions have oxidation state Mn$^{+2}$ and therefore a $d^5$ electronic configuration.  One therefore expects the Mn ions to have high-spin $S = 5/2$ and a high-$T$ molar magnetic entropy of $2R\ln(2S+1)= 2R\ln(6) = 29.79$~J/mol\,K, where $R$ is the molar gas constant and a ``mol'' refers here to a mole of formula units of ${\rm BiMn_2PO_6}$, each of which contains two Mn atoms.  To test this hypothesis, we calculated the magnetic entropy $S_{\rm mag}(T)$ from the $C_{\rm mag}(T)/T$ versus~$T$ data in Fig.~\ref{cp}(b) (blue symbols) according to
\be
S_{\rm mag}(T) = \int_{\rm 3.0\,K}^T\frac{C_{\rm mag}(T^\prime)}{T^\prime}\,dT^\prime,
\ee
where 3.0~K is the low-$T$ limit of the data.  The derived $S_{\rm mag}(T)$ is shown as the red curve in Fig.~\ref{cp}(b).  The value of $S_{\rm mag}$ at 100~K is $(31.7\pm 1.8)$~J/mol\,K, which agrees with the expected value $2R\ln(2S+1) = 29.8$~J/mol\,K for $S=5/2$ within the approximate systematic error bar. Thus we conclude that the Mn$^{+2}$ cations indeed have spin $S = 5/2$.

\begin{figure}[t]
\includegraphics[width=3.in]{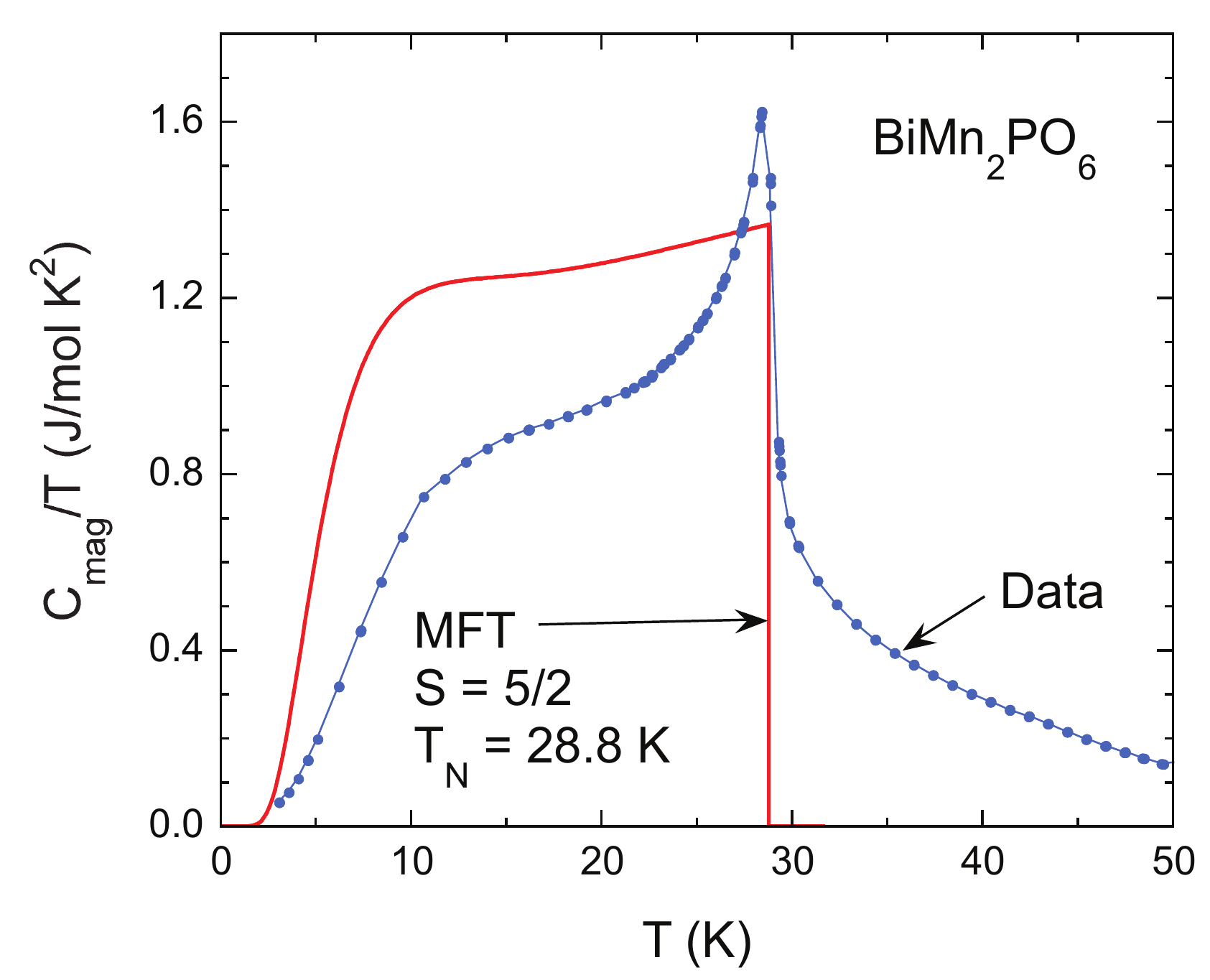}
\caption{(Color online)  Expanded plot of the magnetic heat capacity $C_{\rm mag}(T)/T$ versus $T$ below 50~K (filled blue circles).  The connecting blue line is a guide to the eye.  Also shown is the prediction of molecular field theory (MFT) for spin $S = 5/2$ and N\'eel temperature $T_{\rm N} = 28.8$~K (red curve) which shows a strong broad peak near 10~K\@.  Therefore the broad hump in the experimental data at $\sim 10$~K is due to the magnetic ordering transition at 28.8~K and not to an additional transition at $\sim10$~K\@.}
\label{Fig:BiMn2PO6_CmagOnT}
\end{figure}

An expanded plot of $C_{\rm mag}(T)/T$ versus~$T$ for ${\rm BiMn_2PO_6}$ is shown in Fig.~\ref{Fig:BiMn2PO6_CmagOnT}, together with a fit by the Weiss molecular field theory\cite{johnston2011} (MFT) for spin~$S=5/2$ and N\'eel temperature $T_{\rm N} = 28.8$~K\@.  The broad hump in both the data and MFT at $T\approx 10$~K is due to the combined $T$-dependent influences below $T_{\rm N}$ of the populations of the Zeeman levels and the energies of those levels arising from the $T$-dependent exchange field, which becomes more pronounced as $S$ increases.\cite{johnston2011}  This bulge must increasingly occur with increasing $S$ in order that the entropy at $T_{\rm N}$ increases with increasing $S$, since according to MFT, $C_{\rm mag}(T)$ is bounded from above by the classical prediction.\cite{johnston2011}

According to Eq.~(5), the entropy change over a given $T$ range is the area under the $C_{\rm mag}(T)/T$ versus~$T$ plot over that $T$ range.  Since from Fig.~\ref{cp}(b) the entropy at 100~K of the Mn spins $S = 5/2$ in ${\rm BiMn_2PO_6}$ is completely recovered [$S_{\rm mag} = 2R\ln(2S+1)$], the missing area between the MFT curve and the data in Fig.~\ref{Fig:BiMn2PO6_CmagOnT} for $T < T_{\rm N}$ is recovered at $T > T_{\rm N}$ where the latter entropy gain is due to loss of short-range AFM ordering of the Mn spins with increasing~$T$ above~$T_{\rm N}$. There is no clear evidence in Fig.~\ref{Fig:BiMn2PO6_CmagOnT} for any magnetic transition at about 10~K that was suggested above from the $M(H,T)$ data. Thus this transition does not cause much change in the $T$ dependence of the magnetic entropy of the system.

\begin{figure}[t]
\includegraphics[width=3.in]{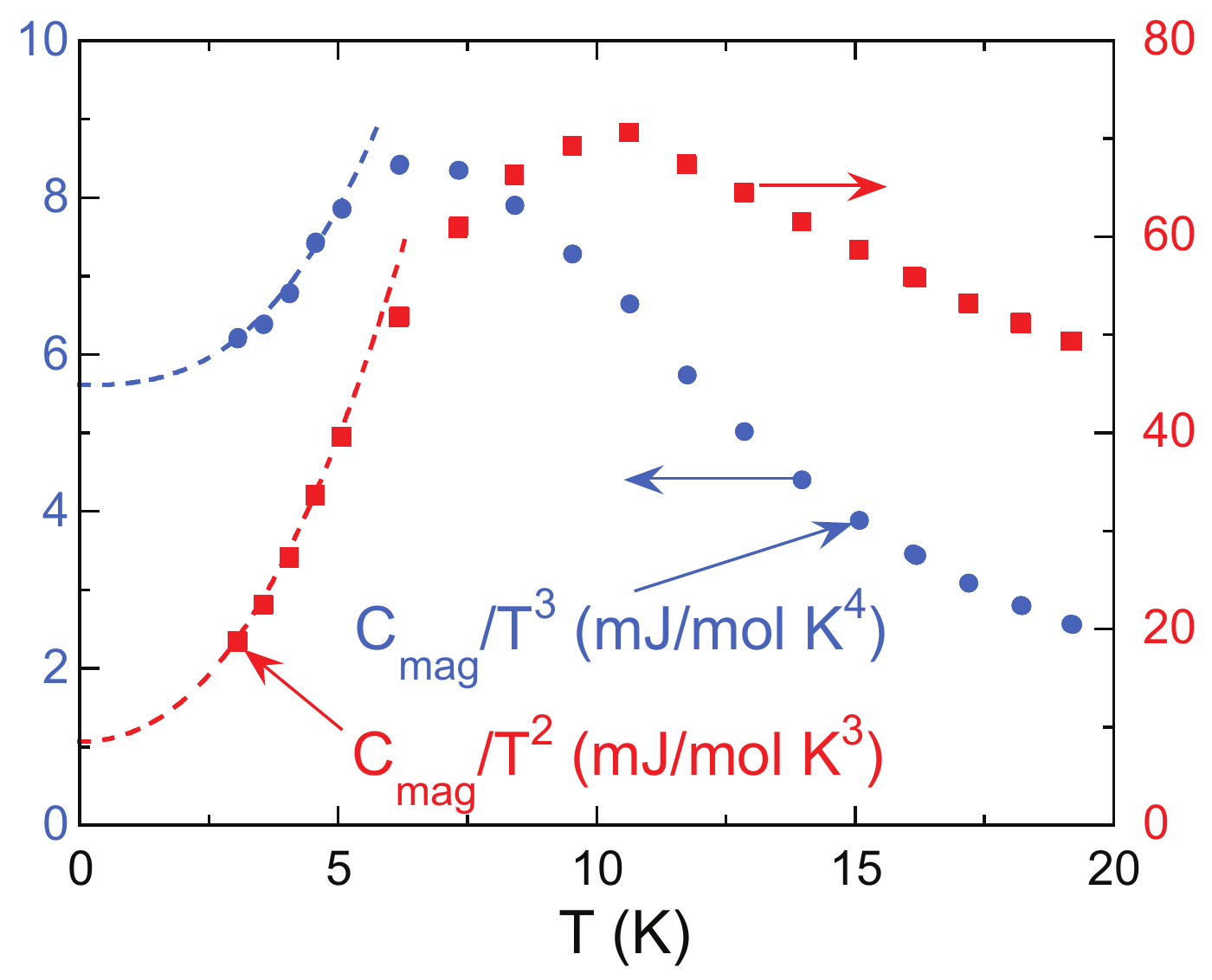}
\caption{(Color online)  Expanded plots of $C_{\rm mag}/T^2$ (red, right ordinate) and $C_{\rm mag}/T^3$ (blue, left ordinate) versus $T$ from Fig.~\ref{Fig:BiMn2PO6_CmagOnT}.  The dashed curves are power-law extrapolations of the data below 5~K to $T=0$ discussed in the text.}
\label{Fig:BiMn2PO6_CmagOnTn}
\end{figure}

The above MFT prediction of $C_{\rm mag}(T)$ for ${\rm BiMn_2PO_6}$ is exponential at low~$T$ because the local exchange field seen by each Mn spin lifts the Zeeman degeneracy which results in energy gaps between the ground and excited Zeeman energy levels of the Mn spin.  On the other hand MFT does not take into account spin-wave excitations in the 3D ordered state which would give rise, in the absence of anisotropy effects leading to an anisotropy gap, to a $T^2$ or $T^3$ dependence of $C_{\rm mag}$ at low~$T$ for spin waves confined mainly to a plane (quasi-2D) or spin waves traveling more or less equally in all three directions (3D), respectively.  Shown in Fig.~\ref{Fig:BiMn2PO6_CmagOnTn} are plots of $C_{\rm mag}/T^2$ (right ordinate) and $C_{\rm mag}/T^3$ (left ordinate) to examine these two possibilities, respectively.  As shown by the dashed-curve power-law extrapolations of the data below 5~K to $T=0$ which both give nonzero intercepts, either case appears to be consistent the data, where the intercepts for $T\to0$ give the potential spin-wave (SW) contributions
\bse
\label{Eqs:CmagSWExpt}
\bea
C_{\rm mag} &=& \beta_{\rm SW} T^3,~~~~({\rm 3D})\\*
\beta_{\rm SW} &\approx& 5.6\,{\rm \frac{mJ}{mol\,K^4}}
\eea
or
\bea
C_{\rm mag} &=& \delta_{\rm SW} T^2,~~~~({\rm 2D})\\*
\delta_{\rm SW} &\approx& 1.0\,{\rm \frac{mJ}{mol\,K^3}}.
\eea
\ese
A quantitative evaluation of the spin-wave contribution to the heat capacity is given below in Sec.~\ref{Sec:SpinWaveCp}.

\subsection{Microscopic Magnetic Model}
\label{sec:dft}
\subsubsection{Evaluation of Magnetic Couplings}

\begin{figure}
\includegraphics{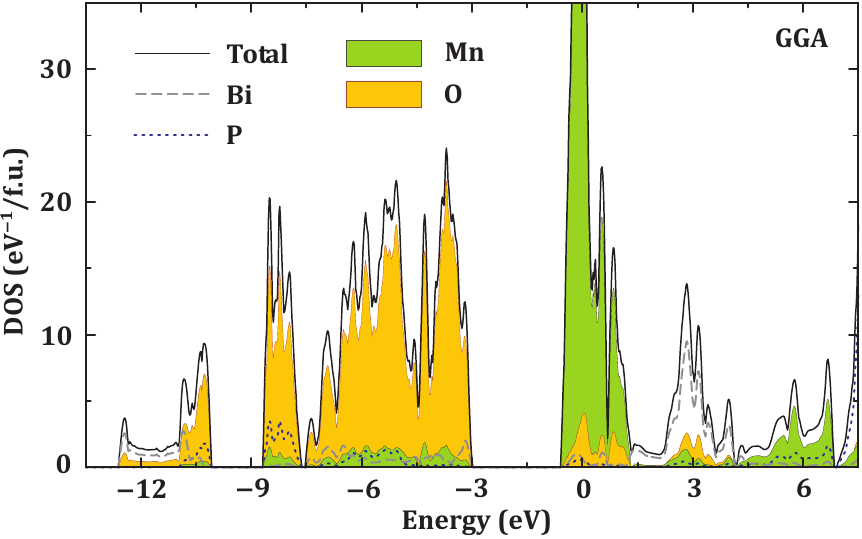}
\caption{\label{fig:dos}
(Color online) GGA electronic density of states (DOS) versus energy for BiMn$_2$PO$_6$. The Fermi energy is defined as zero. Note the nearly isolated $3d$ bands of Mn at energy $\approx 0$ with only a minor contribution of O $2p$ states.}
\end{figure}

The electronic density of states (DOS) versus energy calculated for BiMn$_2$PO$_6$ is shown in Fig.~\ref{fig:dos}. Although BiMn$_2$PO$_6$ is greenish-gray-colored and clearly insulating, we find a metallic DOS, because the calculation is done on the simple GGA level without introducing either the AFM spin polarization or the GGA+$U$ correction for correlation effects, which are both responsible for opening the band gap in an antiferromagnetic Mott insulator. Nevertheless, this simplistic calculation is useful for a direct comparison to the isostructural spin-ladder compound BiCu$_2$PO$_6$ (see Fig.~2 in Ref.~\onlinecite{tsirlin2010}). The difference in the electron count is immediately reflected in the position of the Fermi energy that lies in the middle of the $3d$ band for Mn$^{+2}$ (5 $d$-electrons) and in the top part of the $3d$ band for Cu$^{+2}$ (9 $d$-electrons). Additionally, the nature of states in the vicinity of the Fermi energy is quite different. In BiCu$_2$PO$_6$, 40\,\% of these states are formed by oxygen, whereas in BiMn$_2$PO$_6$ the hybridization with oxygen is much weaker, so that oxygen contributes only 6\% of the states at the Fermi level.

\begin{table}
\caption{\label{tab:exchanges}
Exchange couplings in BiMn$_2$PO$_6$: the Mn--Mn distances (in~\r A), type of the coupling (intra- or interladder), coordination numbers $z_{ij}$ (number of couplings per Mn$^{+2}$ ion), exchange integrals $J_{ij}$ (in~K) defined in Eq.~\eqref{eq:ham}, and normalized spin-spin correlations $\sisj/S^2 \equiv \cos\phi_{ji}$, where $\phi_{ji}$ is the angle between the ordered moments $\vec{\mu}_j$ and $\vec{\mu}_i$ in the ordered AFM state. The exchange integrals are calculated within GGA+$U$ to an accuracy of 0.1\,K using $U_d=5.5$\,eV and $J_d=1$\,eV. The last column lists relevant exchange couplings in BiCu$_2$PO$_6$ according to Ref.~\onlinecite{tsirlin2010}. For the notation of $J_{ij}$, see Fig.~\ref{structure}. The exchange bonds with $\sisj/S^2=+1$ are frustrating for AFM ordering in the proposed structure.}
\begin{ruledtabular}
\begin{tabular}{ccccccc@{\hspace{1em}}c}
 \multicolumn{2}{c}{Bi$M_2$PO$_6$:} &  \multicolumn{4}{c}{$M$=Mn} & $M$=Cu \\
          & $d_{\text{Mn--Mn}}$ & type     &  $z_{ij}$ &    $J_{ij}$    & $\sisj/S^2$ &   $J_{ij}$    \\\hline
$J_3$     &      3.229          & Mn1--Mn2 (intra) &   1   &     5.6     &   $-1.0$                      &   22       \\
$J_1$     &      3.627          & Mn1--Mn2 (intra) &   2   &     6.7     &   $-1.0$                      &  176       \\
$J_{a1}$  &      4.556          & Mn1--Mn2 (inter) &   2   &     0.35    &   $+1.0$                      &  $<5$      \\
$J_{d1}$  &      4.814          & Mn1--Mn1 (intra) &   2   &     0.8     &   $+1.0$                      &  $<5$      \\
$J_{d2}$  &      4.898          & Mn2--Mn2 (intra) &   2   &     0.7     &   $+1.0$                      &  $<5$      \\
$J_4$     &      4.900          & Mn1--Mn2 (inter) &   1   &     2.2     &   $-1.0$                      &  154       \\
$J_2$     &      5.370          & Mn1--Mn1 (intra) &   2   &     0.9     &   $+1.0$                      &  170       \\
$J_2'$    &      5.370          & Mn2--Mn2 (intra) &   2   &     1.3     &   $+1.0$                      &   90       \\
$J_{a2}$  &      6.019          & Mn1--Mn1 (inter) &   4   &     0.6     &   $-1.0$                      &  $<5$      \\
$J_{a2}'$ &      6.078          & Mn2--Mn2 (inter) &   4   &     0.4     &   $-1.0$                      &  $<5$      \\
\end{tabular}
\end{ruledtabular}
\end{table}
Exchange couplings obtained from the supercell GGA+$U$ method are listed in Table~\ref{tab:exchanges}. They enter the following spin Hamiltonian
\begin{equation}
{\cal H}=\sum_{\langle ij\rangle}J_{ij}\mathbf S_i\cdot\mathbf S_j,
\label{eq:ham}\end{equation}
where the summation is over all distinct pairs $\langle ij\rangle$ of Mn atoms, and $\mathbf S_i$, $\mathbf S_j$ are the spin operators for spin-$\frac52$ Mn$^{2+}$ ions. We calculated all interactions for Mn--Mn distances up to 7\,\r A and repeated calculations for different supercells to make sure that longer pathways can be neglected. Error bars in calculated exchange integrals are below 0.1\,K for a given $U_d$ value in GGA+$U$. 

We find that BiMn$_2$PO$_6$ follows the conventional spin-ladder scenario, albeit with a large number of significant interladder couplings. The couplings $J_1$ and $J_3$ along the leg and along the rung of the ladder, respectively, are the two leading interactions in this system (note that we use the notation of Ref.~\onlinecite{tsirlin2010}, which may be a bit counterintuitive here, but facilitates the comparison to BiCu$_2$PO$_6$). These two couplings follow the short Mn--O--Mn pathways and can be analyzed in terms of Goodenough-Kanamori-Anderson (GKA) rules.\cite{goodenough1958} Considering the Mn--O--Mn angles of $116.5^{\circ}$ for $J_1$ and $96.3^{\circ}$ for $J_3$, one may expect a much weaker AFM or even a ferromagnetic (FM) exchange $J_3$, in contrast to the robust AFM exchange $J_1$.
On the other hand, the short \mbox{Mn--Mn} distance between the Mn1O$_5$--Mn2O$_5$ pyramids may facilitate the direct Mn--Mn exchange for $J_3$ and provide an additional source of the AFM coupling, thus leading to a nearly ideal spin ladder with $J_1\simeq J_3$.

According to Table~\ref{tab:exchanges}, both BiMn$_2$PO$_6$ and BiCu$_2$PO$_6$ feature solely AFM exchange, but the couplings between the spin-$\frac52$ Mn$^{+2}$ ions are much weaker than those between spin-$\frac12$ Cu$^{+2}$, as previously seen in the spin-chain compound BaMn$_2$Si$_2$O$_7$ ($J\simeq 12$\,K)\cite{ma2013} versus isostructural BaCu$_2$Si$_2$O$_7$ ($J\simeq 280$\,K).\cite{tsukada1999} This large difference stems from the reduced hybridization between the Mn $3d$ and O $2p$ states that renders superexchange less efficient.

Long-range couplings form triangular loops (Fig.~\ref{structure}) and frustrate the spin lattice of BiMn$_2$PO$_6$. These couplings follow Mn--O$\ldots$O--Mn pathways and remain relatively weak, below 2.5\,K, compared to BiCu$_2$PO$_6$, where the long-range couplings $J_2$, $J_2'$, and $J_4$ are integral to the magnetic model.\cite{tsirlin2010} This difference between the Mn$^{+2}$ and Cu$^{+2}$ compounds should be again traced back to the weaker Mn--O hybridization. 

Altogether, we find that BiMn$_2$PO$_6$ entails stronger couplings along the legs and rungs of the spin ladder and weaker interladder couplings, although the resulting spatial anisotropy is not very strong and does not lead to a truly quasi-1D behavior (see Sec.~\ref{sec:microscopic} for a further discussion). In contrast, BiCu$_2$PO$_6$ is either quasi-1D or quasi-2D and features unexpected long-range couplings along $b$ and $c$ as well as very weak interladder couplings along $a$.

\subsubsection{Molecular Field Theory}
In order to compare calculated exchange couplings with the experiment, we develop Weiss molecular field theory (MFT) for BiMn$_2$PO$_6$. For simplicity, we consider a Heisenberg model with no anisotropy terms except that possibly due to an infinitesimal applied magnetic field {\bf H}\@. The part ${\cal H}_i$ of the spin Hamiltonian associated with a particular central spin ${\bf S}_i$ interacting with its neighbors ${\bf S}_j$ with respective exchange constants $J_{ij}$ is
\be
{\cal H}_i = \frac{1}{2}{\bf S}_{i}\cdot\sum_j J_{ij}{\bf S}_j + g\mu_{\rm B} {\bf S}_i\cdot{\bf H},
\label{Eq:Hamili1}
\ee
where the factor of 1/2 recognizes that the exchange energy is evenly split between  two interacting spins, $g$~is the spectroscopic splitting factor ($g$-factor) of a magnetic moment $\vec{\mu}$, and $\mu_{\rm B}$ is the Bohr magneton. In the Weiss MFT, one only considers the thermal-average directions of  ${\bf S}_i$ and ${\bf S}_j$ when calculating their interaction.  Furthermore, it is the magnetic moment $\vec{\mu}$ that interacts with a magnetic field and not the angular momentum {\bf S} \emph{per~se}.  The relationship between these two quantities for an electronic spin and magnetic moment is
\be
{\bf S} = -\frac{\vec{\mu}}{g\mu_{\rm B}},
\label{Eq:Sfrommu}
\ee
where the minus sign arises from the negative sign of the electron charge.  In the following, the symbol $\vec{\mu}$ refers to the thermal-average value of a magnetic moment, as is appropriate in MFT\@. Then the energy $E_i$ of interaction of magnetic moment $\vec{\mu}_i$ with its neighbors $\vec{\mu}_j$ is given by Eq.~(\ref{Eq:Hamili1}) as
\be
E_i = \frac{1}{2g^2\mu_{\rm B}^2}\vec{\mu}_i\cdot\bigg(\sum_j J_{ij}\vec{\mu}_j\bigg) - \vec{\mu}_i \cdot{\bf H}.
\label{Eq:Hamili2}
\ee

In MFT, one replaces the sum of the exchange interactions acting on $\vec{\mu}_i$ in the first term by an effective magnetic field called the Weiss molecular field ${\bf H}_{\rm exch}$, or ``exchange field'', that is defined by the usual relationship for the rotational potential energy of a magnetic moment in a magnetic field, as in the second term of Eq.~(\ref{Eq:Hamili2}), as
\be
2E_{{\rm exch}\,i} = -\vec{\mu}_i\cdot{\bf H}_{\rm exch},
\label{Eq:EexchtoHexch}
\ee
 where the factor of 2 arises because in MFT all of the exchange energy between $\vec{\mu}_i$ and $\vec{\mu}_j$ is attributed to $\vec{\mu}_j$, thus canceling out the factor of 1/2 in Eq.~(\ref{Eq:Hamili2}).  From the first term in Eq.~(\ref{Eq:Hamili2}) one obtains
\be
{\bf H}_{{\rm exch}\,i} = -\frac{1}{g^2\mu_{\rm B}^2}\sum_j J_{ij} \vec{\mu}_{j}.\\*
\label{Eq:HexchiDef}
\ee
Using $\vec{\mu}_{j} = \mu_j\hat{\mu}_j$ where $\mu_j = |\vec{\mu}_{j}|$, the component of ${\bf H}_{{\rm exch}\,i}$ in the direction of $\vec{\mu}_{i}$ is
\bea
H_{{\rm exch}\,i} &=& \hat{\mu}_i\cdot {\bf H}_{{\rm exch}\,i} = -\frac{1}{g^2\mu_{\rm B}^2}\sum_j J_{ij}\mu_j\hat{\mu}_i\cdot\hat{\mu}_j\nonumber\\*
&=& -\frac{1}{g^2\mu_{\rm B}^2}\sum_j J_{ij}\mu_j\cos\phi_{ji},\label{Eq:HexchDef3}
\eea
where $\phi_{ji}$ is the angle between $\vec{\mu}_{j}$ and $\vec{\mu}_{i}$.

Now we specialize the treatment to a local-moment magnetic system containing two crystallographically inequivalent sublattices~1 and~2 of identical spins as occurs in \bi\ with the presence of the Mn1 and Mn2 spins-5/2, respectively.  One can separate the sum in Eq.~(\ref{Eq:HexchDef3}) into two sums over spins in the same~(s) and different~(d) sublattices~1 and~2 of Mn1 and Mn2, yielding
\bse
\label{Eqs:Hexch12i}
\be
H_{{\rm exch}\,1i} = -\frac{1}{g^2\mu_{\rm B}^2}\bigg({\sum_j}^{\rm s} J_{ij}\mu_{1j}\cos\phi_{ji} + {\sum_j}^{\rm d}J_{ij}\mu_{2j}\cos\phi_{ji}\bigg),
\label{Eq:HexchDefMn1}
\ee
\be
H_{{\rm exch}\,2i} = -\frac{1}{g^2\mu_{\rm B}^2}\bigg({\sum_j}^{\rm d} J_{ij}\mu_{1j}\cos\phi_{ji} + {\sum_j}^{\rm s}J_{ij}\mu_{2j}\cos\phi_{ji}\bigg),
\label{Eq:HexchDefMn2}
\ee
\ese
where $H_{{\rm exch}\,1i}$ is the exchange field seen by a Mn spin on the Mn1 sublattice, $H_{{\rm exch}\,2i}$ is the exchange field seen by a Mn spin on the Mn2 sublattice and $\phi_{ji}$ is the angle between the respective magnetic moments. In the paramagnetic state $\phi_{ji} = 0$ for all spin pairs, since all moments point in the direction of the applied field, whereas in the AFM-ordered state with $H=0$  one has either $\phi_{ji} = 0$ or~$180^\circ$ according to the AFM structure in Table~III\@ deduced from our electronic structure calculations for BiMn$_2$PO$_6$.

In MFT, the response of a given magnetic moment to the exchange and applied fields is governed by the Brillouin function $B_S(y)$ according to
\bse
\label{Eq:BS(y)}
\be
 \mu_i = \mu_{\rm sat}B_S(y_i)
\label{Eq:muvsBrill}
\ee
where
\be
y_i = \frac{g\mu_{\rm B}B_i}{k_{\rm B}T},
\label{Eq:yDef}
\ee
$k_{\rm B}$ is the Boltzmann constant, the component of the local magnetic induction in the direction of $\vec{\mu}_i$ is
\be
B_i = H_{{\rm exch}\,i} + H_i,
\ee
and the saturation moment of each spin is
\be
\mu_{\rm sat} = gS\mu_{\rm B}.
\label{Eq:musat}
\ee
\ese
We write the Brillouin function as
\begin{subequations}
\label{Eqs:BS}
\be
B_S(y) = \frac{1}{2S} \left\{(2S+1)\coth\left[(2S+1)\frac{y}{2}\right]-\coth\left(\frac{y}{2}\right)\right\}
\label{Eq:BrillouinFunction},
\ee
for which the Taylor expansion about $y=0$ is
\be
B_S(y) = \frac{(S+1)y}{3} + {\cal O}(y^3).
\label{Eq:BSyTaylor}
\ee
\end{subequations}

\paragraph{Paramagnetic State.}

In the paramagnetic state all induced moments are lined up with the applied magnetic field and one therefore has $\phi_{ji}=0$ for all spin pairs, we assume infinitesimal~$H$ and therefore $y_i\ll1$, and for a spin in either of the two Mn1 or Mn2 sublattices the above equations then yield
\bea
\mu_i &=& \frac{g^2S(S+1)\mu_{\rm B}^2}{3k_{\rm B}T}(H_{{\rm exch}\,i} + H)\label{Eq:mui}\\*
&=& \frac{C_1}{T}\bigg[-\frac{\mu_i}{g^2\mu_{\rm B}^2}\Big({\sum_j}^{\rm s} J_{ij} + {\sum_j}^{\rm d}J_{ij}\Big) + H\bigg],\nonumber
\eea
where the single-spin Curie constant is
\be
C_1 = \frac{g^2S(S+1)\mu_{\rm B}^2}{3k_{\rm B}}.
\ee
Solving Eq.~(\ref{Eq:mui}) for $\mu_i$ gives the Curie-Weiss law
\be
\mu_i = \frac{C_1H}{T-\theta_{\rm CW}},
\ee
where the Weiss temperature is
\be
\theta_{\rm CW} = -\frac{S(S+1)}{3k_{\rm B}}\Big({\sum_j}^{\rm d} J_{ij} + {\sum_j}^{\rm s}J_{ij}\Big).
\label{Eq:CWtheta}
\ee
This treatment is valid for \bi\ because Table~III gives, for both Mn sublattices in \bi, the similar values
\be
{\sum_j}^{\rm d} J_{ij}/k_{\rm B} = 21.9~{\rm K}, \quad {\sum_j}^{\rm s} J_{ij}/k_{\rm B} = 5.7(1)~{\rm K},
\label{Eq:JSums}
\ee
where the error bar on the second sum reflects the difference between the sums obtained for Mn1 and Mn2 as the central spin~$i$ obtained from Table~\ref{tab:exchanges}\@.  This error bar does not include the error bar of 0.1\,K in the calculation of the $J_{ij}$ values themselves.  Thus for the Mn spins $S=5/2 $ in \bi, Eq.~(\ref{Eq:CWtheta}) predicts the Weiss temperature to be
\be
\theta_{\rm CW}^{\rm calc} = -80.5(3)~{\rm K},
\ee
which compares favorably with the value of $-78$~K obtained from the fit of the experimental data by the Curie-Weiss law in Fig.~\ref{chi}(a).

\paragraph{Antiferromagnetic State.}

Within MFT, we obtain the N\'eel temperature~$T_{\rm N}$ by setting the magnitudes of the ordered moments of all the Mn spins to be the same, $\mu_{1j} = \mu_{2j} \equiv \mu_i$, using the values of $\phi_{ji}$ in Eqs.~(\ref{Eqs:Hexch12i}) as given in Table~III, canceling out the factor of $\mu_i\to0$ for $T\to T_{\rm N}^-$ on both sides of Eq.~(\ref{Eq:muvsBrill}) using the expansion~(\ref{Eq:BSyTaylor}), and solving for $T\equiv T_{\rm N}$, yielding
\be
T_{\rm N} = -\frac{S(S+1)}{3k_{\rm B}}\Big({\sum_j}^{\rm d}J_{ij}\cos\phi_{ji} + {\sum_j}^{\rm s}J_{ij}\cos\phi_{ji}\Big).
\label{Eq:TNcalc}
\ee
The values of the sums are obtained from the $J_{ij}$ data and the $\cos\phi_{ji}$ values for the calculated AFM structure in Table~III\@.  The reason that the $\cos\phi_{ji}$ factor is included even in the second sum over Mn spins on the same sublattice is that some of these spins are parallel to a given Mn spin on this sublattice and some are antiparallel according to Table~III\@ and Fig.~\ref{structure}. Using the data in Table~III, Eq.~(\ref{Eq:TNcalc}) yields
\bse
\label{Eqs:JSums2}
\bea
{\sum_j}^{\rm d} (J_{ij}/k_{\rm B})\cos\phi_{ji} &=& -20.5~{\rm K}, \\*
{\sum_j}^{\rm s} (J_{ij}/k_{\rm B})\cos\phi_{ji} &=& 1.7(7)~{\rm K},
\eea
\ese
where the error bar on the second sum again reflects the difference between the sums obtained for Mn1 and Mn2 as the central spin~$i$.  Using the values of the sums in Eqs.~(\ref{Eqs:JSums2}) gives the prediction of MFT for $T_{\rm N}$ from Eq.~(\ref{Eq:TNcalc}) as
\be
T_{\rm N}^{\rm calc} = 55(2)~{\rm K}.
\label{Eq:TNcalc2}
\ee
This predicted value of $T_{\rm N}$ is about a factor of two larger than the observed value $T_{\rm N} \approx 30$~K\@.

Using Eqs.~(\ref{Eq:CWtheta}) and~(\ref{Eq:TNcalc2}) one obtains the calculated frustration parameter
\be
f^{\rm calc}\equiv\frac{|\theta_{\rm CW}^{\rm calc}|}{T_{\rm N}^{\rm calc}} = \frac{{\sum_j}^{\rm d} J_{ij} + {\sum_j}^{\rm s}J_{ij}}{{\sum_j}^{\rm d}J_{ij}\cos\phi_{ji} + {\sum_j}^{\rm s}J_{ij}\cos\phi_{ji}}.
\ee
The above values for $\theta_{\rm CW}^{\rm calc}$ and~$T_{\rm N}^{\rm calc}$ then yield
\be
f^{\rm calc} = 1.52(7),
\ee
which is significantly smaller than the observed value of about 2.6.  This suppression of~$f$ is due to neglect by MFT of the influences of quantum fluctuations associated with frustration for AFM ordering and spatial anisotropy of the Mn--Mn exchange interactions as discussed below in Sec.~\ref{sec:microscopic}.  These two factors together suppress the observed $T_{\rm N}\approx 30$~K to be below the MFT estimate $T_{\rm N}^{\rm calc} = 55~{\rm K}$ in Eq.~(\ref{Eq:TNcalc2}).  This suppression of $T_{\rm N}$ below the MFT value leads to the observed value of $f$ being larger than the one calculated using MFT\@.

\subsubsection{\label{Sec:CMC} Monte-Carlo Simulations}

\begin{figure}
\includegraphics{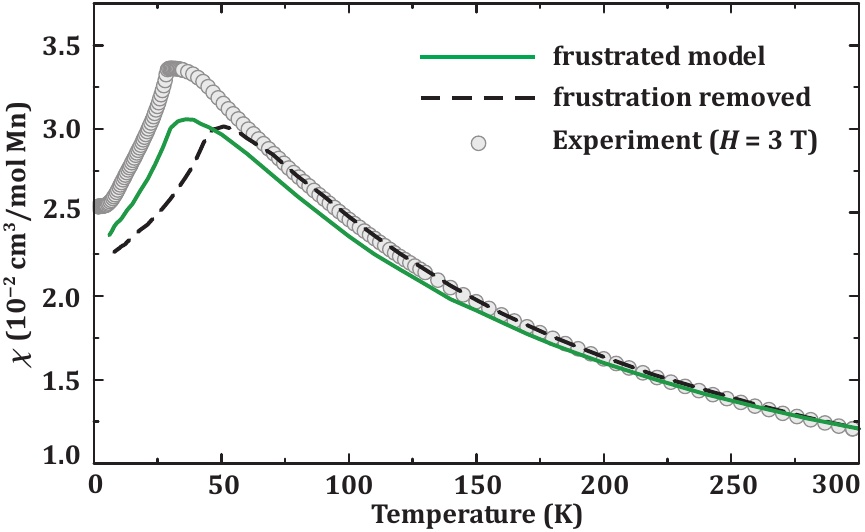}
\caption{\label{fig:fit}
(Color online) Experimental magnetic susceptibility $\chi$ versus temperature measured in an applied field of 3\,T (circles) and the $\chi$ simulated for the frustrated (solid curve, all couplings from Table~\ref{tab:exchanges}) and nonfrustrated (dashed curve, the frustrating couplings in Table~\ref{tab:exchanges} removed) magnetic models of BiMn$_2$PO$_6$.  From the temperatures at which the maximum slope of $\chi(T)$ occur below the $\chi(T)$ maxima, the simulations for the frustrated and nonfrustrated models give AFM ordering temperatures $T_{\rm N} = 27$ and~47~K, respectively.}
\end{figure}

We can also treat the problem numerically by simulating the magnetic susceptibility of our microscopic model with the exchange couplings from Table~\ref{tab:exchanges}. In Fig.~\ref{fig:fit}, we compare a classical Monte-Carlo simulation of the magnetic spin susceptibility with the experimental data collected at 3\,T, where the spurious 43\,K feature is fully suppressed. The simulated spin susceptibility curve has been scaled with $g=2.0$ and shifted by a temperature-independent term $\chi_0=4\times10^{-4}$\,cm$^3$/mol~Mn, according to the Curie-Weiss fit in Sec.~\ref{sec:magnetization}. The overall shape of the experimental curve is reproduced, and the simulated $T_{\rm N}\simeq 27$\,K is in good agreement with the experimental value of about 30~K\@.  However, the absolute values of the susceptibility below 200\,K are slightly underestimated.  This discrepancy requires further investigation. It may be related to the pronounced lattice softening that would modify the exchange couplings $J_{ij}$ (our values in Table~\ref{tab:exchanges} are for the room-temperature crystal structure).

Surprisingly, frustration has no visible effect on the classical ground state of BiMn$_2$PO$_6$. We analyze this ground state by calculating normalized spin-spin correlations $\langle{\bf S}_i\cdot{\bf S}_j\rangle/S^2$ at $T=0.1$\,K using DFT\@. The normalized spin-spin correlation is equal to $+1$ for the parallel spin alignment, $-1$ for the antiparallel spin alignment, and takes intermediate values between $-1$ and $+1$ for noncollinear spin configurations. In our case, all correlations are found to be equal to $\pm1$, hence a collinear long-range order is expected. The ordering pattern is determined by the strongest couplings on each triangular loop. The antiparallel spin arrangement within the ladder is imposed by $J_1$ and $J_3$, the AFM order along $c$ is driven by $J_4$, and the order along $a$ relies on $J_{a2},J_{a2}'>J_{a1}$ (see Fig.~\ref{structure}).

Our experimental data give strong evidence for the magnetic frustration in BiMn$_2$PO$_6$. The experimental ratio $f = |\theta_{\rm CW}|/T_{\rm N} = 2.6$ indicates a moderate magnetic frustration.\cite{ramirez2001} To verify this, we constructed a simplified magnetic model, where the frustration is eliminated by removing the frustrating couplings in Table~\ref{tab:exchanges} for which $\sisj/S^2=-1$, so that only $J_1$, $J_3$, $J_4$, $J_{a2}$, and $J_{a2}'$ remain. From a classical Monte-Carlo simulation of this non-frustrated model, we obtain a much higher N\'eel temperature of about 47\,K (see the dashed curve in Fig.~\ref{fig:fit}), compared to 27~K with the frustrating interactions present.  The corresponding values obtained using MFT are $T_{\rm N} = 55$~K from Eq.~(\ref{Eq:TNcalc2}) with the frustrating interactions included and~$T_{\rm N} = 68$~K without them.  The difference between the two $T_{\rm N}$ values without and with the frustrating interactions present is 20~and~13~K, respectively. Hence MFT underestimates the suppression of $T_{\rm N}$ due to the frustration because it neglects fluctuations associated with it. 
%These fluctuations would also be expected to reduce the ordered (saturation) moment measured by neutron diffraction at low temperatures to a value less than the nominal value $\mu_{\rm sat} = gS\mu_{\rm B}\simeq 5\,\mu_{\rm B}$.\cite{johnston2011}

\subsection{\textbf{$^{31}$P NMR}}

\begin{figure}
\includegraphics [width=3in] {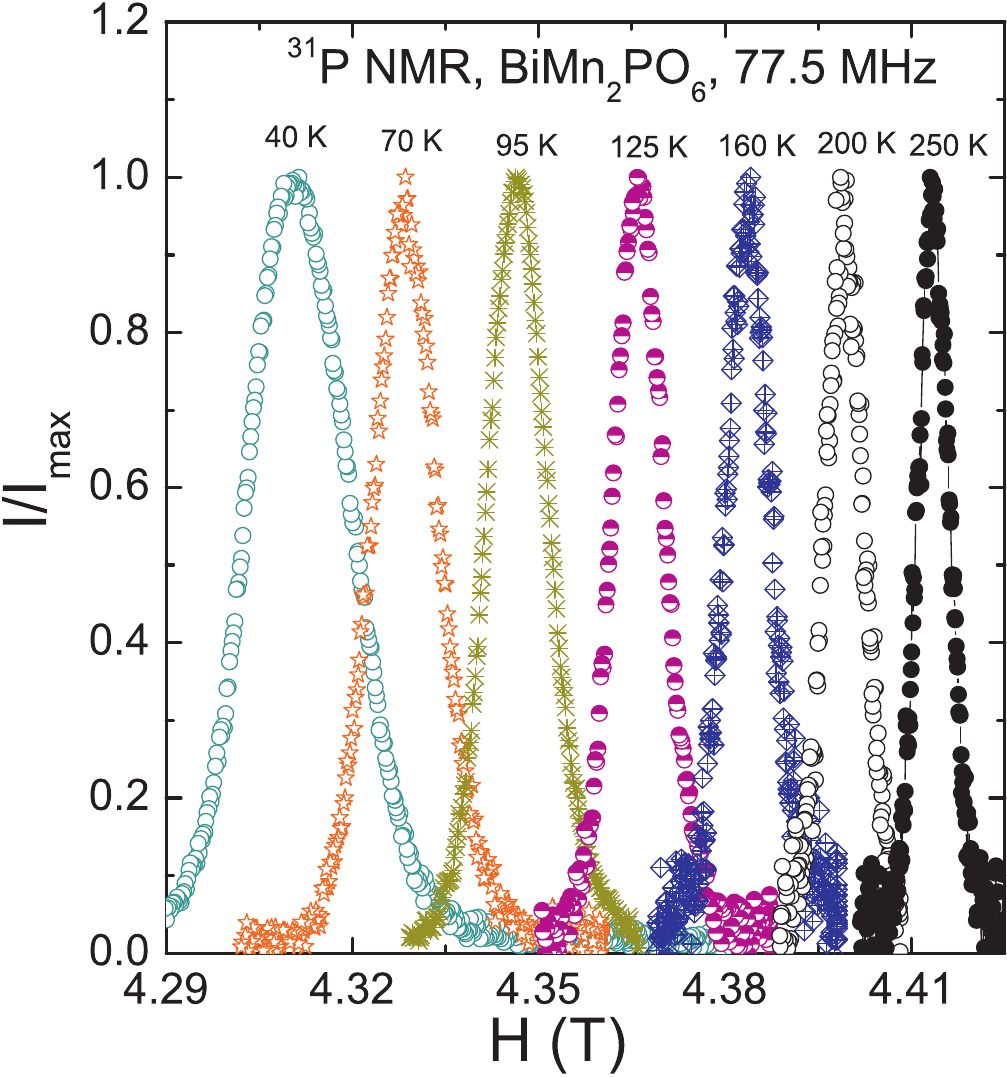}
\caption{\label{spk} (Color online) $^{31}$P NMR spectra of the intensity $I$ versus magnetic field~$H$ measured at 77.5~MHz and at the different temperatures indicated.}
\end{figure}

\begin{figure}
\includegraphics [width=3in] {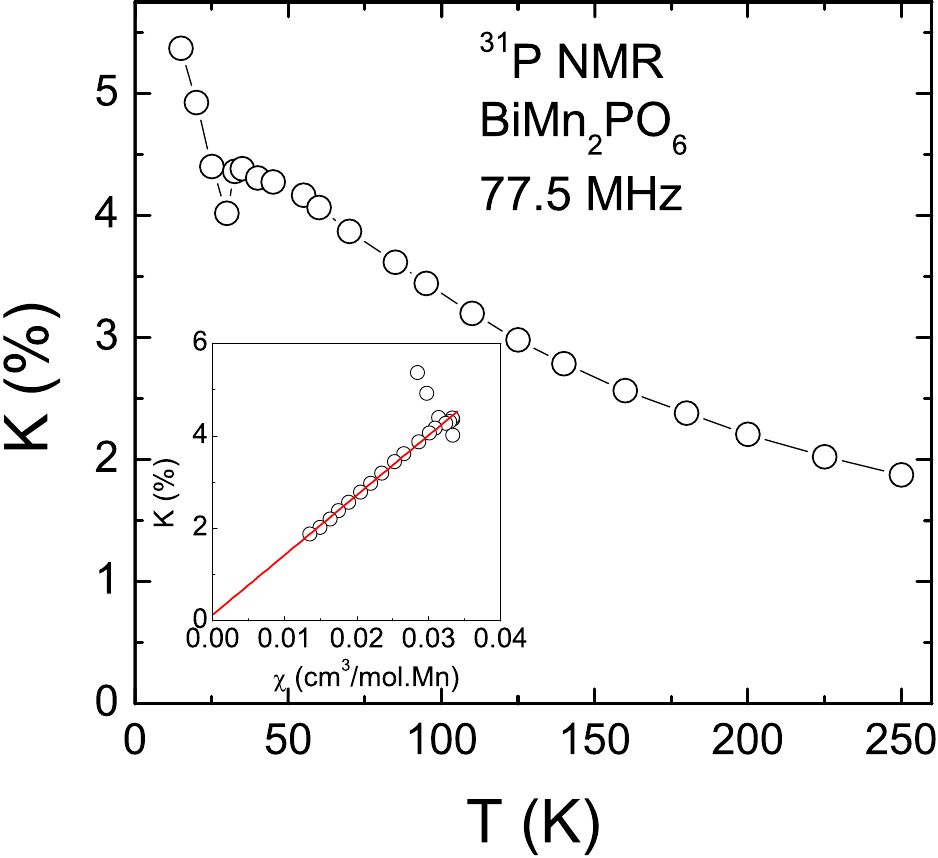}
\caption{\label{K} (Color online) $^{31}$P NMR shift
$K$ versus temperature $T$\@.  Inset: $K$ versus $\chi$ measured at an NMR field of 5~T with temperature as an implicit parameter. The solid red line is the
linear fit.}
\end{figure}
To further study the nature of the magnetic transitions and to elucidate static as well as dynamic properties of BiMn$_2$PO$_6$, we performed $^{31}$P NMR measurements on BiMn$_{2}$PO$_{6}$. An advantage of NMR is that it is not sensitive to impurities. Therefore, one can probe the intrinsic properties of the system. Since all P atoms are crystallographically equivalent (see Table~\ref{Cry_parameters1}),\cite{xun2002} for a spin $I=\frac12$ nucleus one would expect a single spectral line.\cite{nath2005,nath2008b} Indeed, we observe one narrow spectral line. Figure~\ref{spk} shows the $^{31}$P NMR spectra measured at different temperatures. The line position was found to shift with temperature. Figure~\ref{K} presents the $T$-dependence of the NMR shift, $K(T)$. At high-$T$, $K$ varies in a Curie-Weiss manner and shows a change in slope at about 30\,K associated with the AFM ordering.

Since the NMR shift is a direct measure of the spin susceptibility $\chi_{\rm spin}$, one can write $K (T)$ in terms of $\chi_{\rm spin}(T)$ as
\begin{equation}
K(T)=K_{0}+\frac{A_{\rm hf}}{N_{\rm A}} \chi_{\rm spin}(T),
\label{shift}
\end{equation}
where $K_{0}$ is the $T$-independent chemical shift, $A_{\rm hf}$ is the hyperfine coupling constant
of the $^{31}$P nuclei to the Mn$^{+2}$ spins and $N_{\rm A}$ is Avogadro's number. The conventional scheme for calculating
$A_{\rm hf}$ is to obtain it from the slope of a $K$ versus $\chi$ plot with $T$ as an
implicit parameter.  As seen in the inset of Fig.~\ref{K}, the $K$ versus $\chi$ plot
is a nice straight line at high temperatures ($T=35-250$\,K) yielding $K_0=(0.13 \pm 0.03)$\% and $A_{\rm hf} =
(7224 \pm 85)$\,Oe/$\mu_{\rm B}$. The temperature-independent shift $K_0$ contains an intrinsic chemical shift together with extrinsic contributions, including the remnant field of the field-sweep magnet that is not known exactly.

The total hyperfine coupling constant
at the P site is generally the sum of the transferred hyperfine ($A_{\rm trans}$) and
dipolar ($A_{\rm dip}$) couplings produced by the Mn$^{+2}$ spins, i.e.,
$A_{\rm hf}=z'A_{\rm trans}+A_{\rm dip}$, where $z'$ is the number of nearest-neighbor Mn$^{+2}$
spins of the P-site. The anisotropic dipolar couplings were calculated for three different
orientations using lattice sums. The maximum dipolar field contribution
was calculated to be 800\,Oe/$\mu_{\rm B}$, which is one order of magnitude smaller
than the total hyperfine field, suggesting that the dominant contribution to the
total hyperfine coupling is due to the transferred hyperfine coupling at the P-site.
The total $A_{\rm hf}$ of the P site with the Mn$^{+2}$ ions is 7224~Oe/$\mu_{\rm B}$.
As discussed later, each P atom has $z'=6$ neighboring Mn$^{+2}$ spins, so the $A_{\rm hf}$ due to one spin is $A_{\rm hf}/z'=1.2$\,kOe/($\mu_{\rm B}$~Mn) assuming a uniform hyperfine coupling to all $z'$ Mn spins.

\begin{figure}
\includegraphics [width=3in] {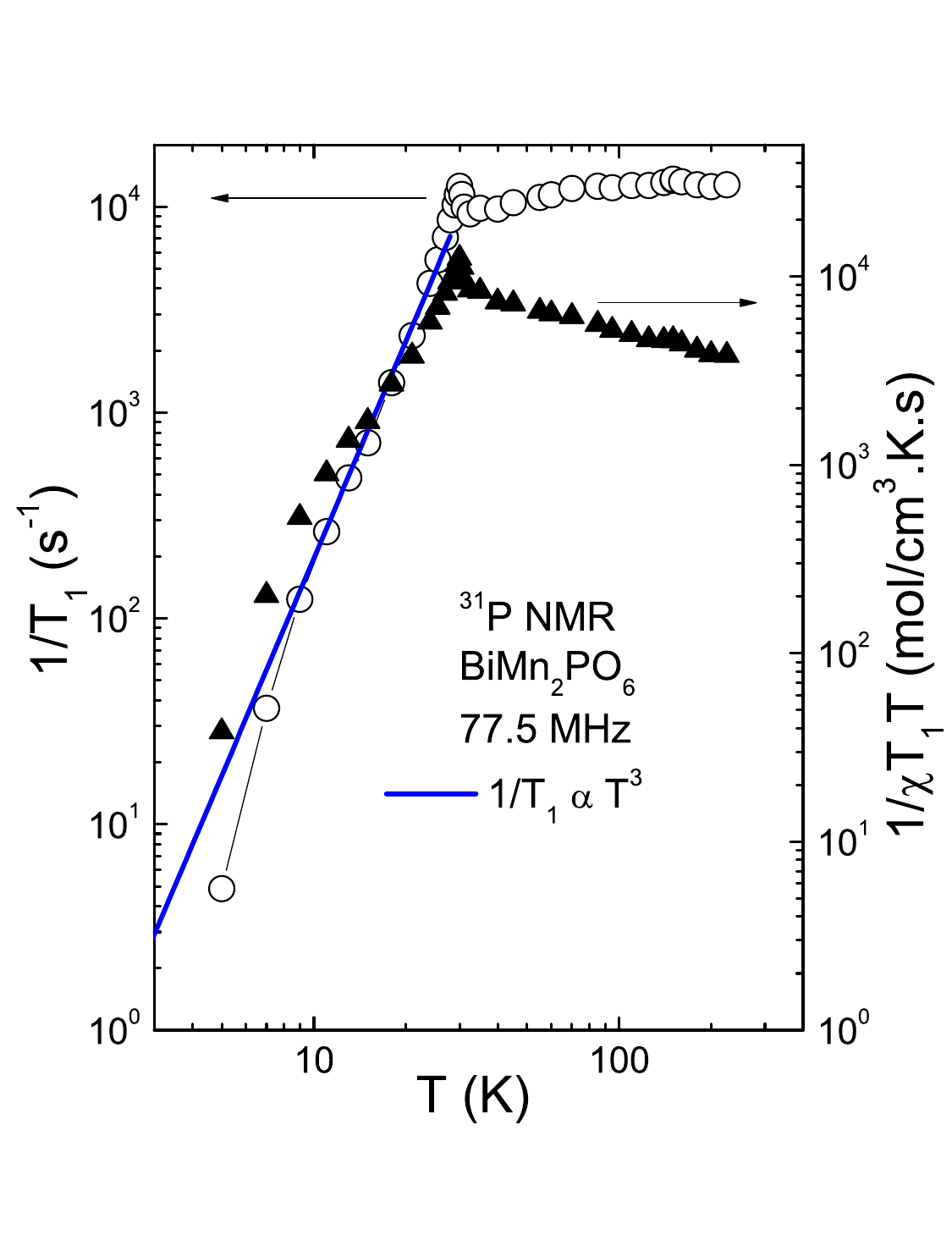}
\caption{\label{t1} (Color online) The $^{31}$P nuclear spin-lattice relaxation rate $1/T_{1}$ and the corresponding $1/\chi T_{1}T$ as a function of temperature $T$ are plotted along the left and right $y$-axes, respectively. The solid blue line corresponds to $1/T_{1} \propto T^{3}$.}
\end{figure}

For an $I=\frac12$ nucleus, the recovery of the longitudinal magnetization
is expected to follow a single-exponential behavior. In BiMn$_{2}$PO$_{6}$,
the recovery of the nuclear magnetization after a comb of saturation pulses was
indeed fitted well by the exponential function $1-\frac{M(t)}{M_{0}}=A\,e^{-t/T_{1}}$,
where $M(t)$ is the nuclear magnetization at time $t$ after
the saturation pulse and $M_{0}$ is the equilibrium magnetization.
The temperature dependence of the nuclear spin-lattice relaxation rate $1/T_{1}$ estimated from the above
fit is presented in Fig.~\ref{t1}.  At high temperatures ($T \gtrsim 70$\,K),
$1/T_{1}$ is almost temperature-independent, which is typical in the
paramagnetic regime ($T\!\gg\! J_{\rm max}/k_{\rm B}$), where $J_{\rm max}$ is the maximum exchange constant in the system.\cite{moriya1956a} With decrease
in $T$, $1/T_{1}$ decreases slowly for $T<70$\,K and then shows a peak at
around 30\,K\@. This decrease in $1/T_{1}$ with decreasing $T$ above $T_{\rm N}$ resembles the behavior of the AFM square-lattice
compound Pb$_{2}$VO(PO$_{4}$)$_{2}$.\cite{nath2009} The peak at $T_{\rm N} \simeq 30$\,K is associated with
the onset of 3D-LRO and is consistent with the thermodynamic
measurements. For $T<T_{\rm N}$, $1/T_{1}$ decreases rapidly.

\begin{figure}
\includegraphics [width=3.3in] {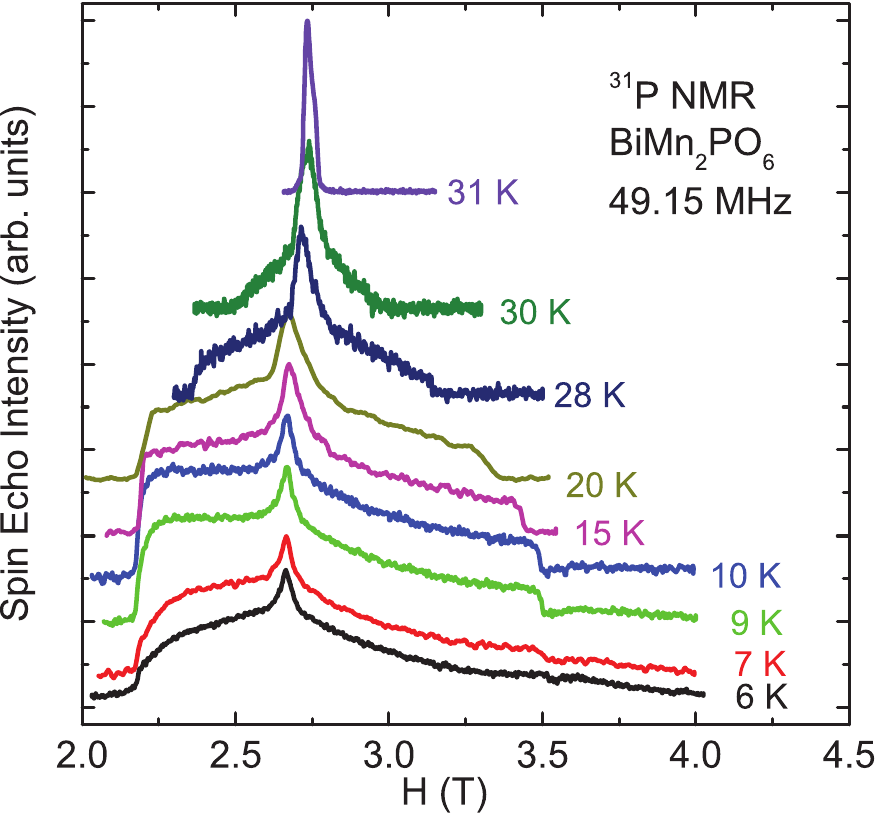}
\caption{\label{lowTspk} (Color online) Field-sweep $^{31}$P NMR spectra
measured at 49.15~MHz in the low-$T$ regime.}
\end{figure}

The $^{31}$P spectrum measured at 77.5\,MHz is broadened
abruptly below $T_{\rm N}$ indicating that the P site is experiencing
the static internal field in the ordered state. In order to precisely
probe the line shape associated with the magnetic ordering, we remeasured
the spectra below 45\,K at a lower frequency of 49.15\,MHz. No noticeable
line broadening was observed around 43\,K, again suggesting that the transition
at 43\,K observed in the above $\chi(T)$ data is extrinsic. As demonstrated in Fig.~\ref{lowTspk},
with decrease in $T$, a systematic line broadening on either side of the
narrow central line occurs below 30\,K\@. This line broadening increases
and the intensity of the central line decreases with decreasing temperature.
At low temperatures the broad line takes almost a rectangular shape down to 10\,K, whereas below 10\,K the edges of the line are smeared following the 10~K magnetic transition which is seen in the $\chi(T)$ measurements in Figs.~\ref{chi}(b) and~\ref{chi}(c). The possible origin of these changes in the line shape is discussed in Sec.~\ref{sec:discussion}A.

Even far below 10~K, the central line related to the high-$T$ paramagnetic
phase does not disappear from the experimental spectra completely. The coexistence
of the high-$T$ phase with the low-$T$ phase has been observed
before in BaCuP$_2$O$_7$,\cite{nath2005}
(Li,Na)VGe$_2$O$_6$,\cite{gavilano2000,vonlanthen2002,pedrini2004} 
and (Ca$_4$Al$_2$O$_6$)Fe$_2$(As$_{1-x}$P$_x$)$_2$.\cite{kinouchi2013} 
One could argue that the coexistence of the two phases is due
to a spread of the transition temperatures within the polycrystalline sample, but in
such a case it would seem quite unlikely to observe the distinct peak in
the temperature dependence of $1/T_1$ as seen in Fig.~\ref{t1}. Another
possible origin of the narrow central line is the presence of crystal
defects or local dislocations in a polycrystalline sample.

A very broad background signal was also observed at 4.2\,K extending over a large field
range. This signal can be attributed to the $^{209}$Bi nuclei.

\section{Discussion}
\label{sec:discussion}

\subsection{Long-range Magnetic Order}

Our thermodynamic and NMR measurements consistently show two intrinsic magnetic transitions in BiMn$_2$PO$_6$. The first transition at $T_{\rm N}\simeq 30$\,K corresponds to the onset of long-range AFM order that manifests itself by the kink of the magnetic susceptibility, the $\lambda$-type anomaly in the specific heat, the maximum in $1/T_1$, and the broadening of the $^{31}$P NMR line. The second transition around 10\,K reveals weaker features reminiscent of a spin reorientation transition. In the following, we analyze experimental signatures of these transitions in NMR.

At $T\!\leq\! T_{\rm N}$, the $^{31}$P NMR line broadens abruptly and has an almost rectangular shape at low temperatures, similar to that reported for (Li,Na)VGe$_{2}$O$_{6}$, CuV$_2$O$_6$, BaCo$_2$V$_2$O$_8$, and BaCuP$_{2}$O$_{7}$ in the AFM-ordered state.\cite{gavilano2000,vonlanthen2002,pedrini2004,kikuchi2000,ideta2012,nath2005} The broad and rectangular NMR spectra at $T \leq T_{\rm N}$ represent the powder spectra of a commensurate antiferromagnetically ordered phase in which the P-site feels the internal field of Mn$^{+2}$ spins.\cite{yamada1986} If the $^{31}$P site is located symmetrically between the neighboring up and down spins, their hyperfine fields induced at this site will be equal and opposite. In this case, one finds a symmetric powder spectra or, for a single crystal, two narrow lines of equal intensity will appear on both sides of the zero-shift position, as in Pb$_2$VO(PO$_4$)$_2$ and (Ba,Sr)Fe$_2$As$_2$.\cite{nath2009,kitagawa2008,kitagawa2009}

\begin{figure}
\includegraphics [width=3.3in] {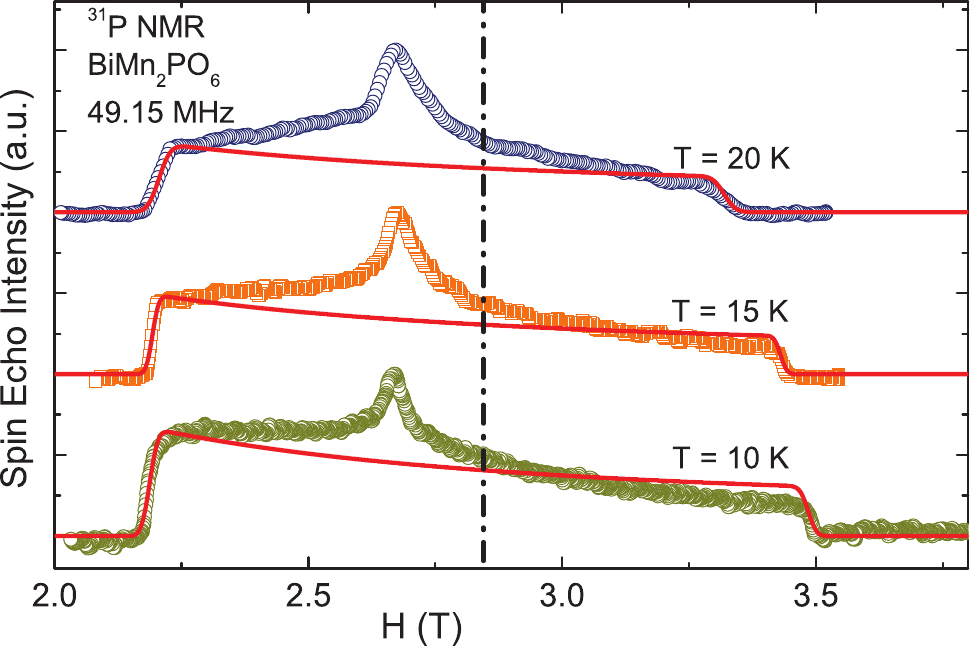}
\caption{\label{spkfit} (Color online) $^{31}$P NMR spectra in the
ordered state at $T<T_{\rm N}\simeq 30$~K measured at 49.15\,MHz. The solid lines represent the
calculated spectra at different temperatures using Eq.~(\ref{convo})
with a distribution function $g(H)=\frac{1}{\sqrt{2\pi\Delta H_{\rm i}^2}}
 \exp \left[-\frac{1}{2} \frac{(H-H_{\rm i})^2}{\Delta H_{\rm i}^2}\right]$.
The vertical dashed line represents the zero-shift central position
$\omega/\gamma_N = 2.845$\,T for $^{31}$P nuclei. The parameters
used to simulate the spectrum at $T=10$\,K are $H_{\rm i}=6.546$\,kOe and
$\Delta H_{\rm i} \simeq 0.13$\,kOe.}
\end{figure}

In order to determine the magnitude of internal field $H_{\rm i}$ at the $^{31}$P NMR site, we calculated the line shape of the NMR spectrum in the AFM ordered state, and fitted the calculated spectrum to the experiment. In powder samples, the angle between the direction of the external magnetic field $H$ and that of internal magnetic field $H_{\rm i}$ due to the AFM ordered spins is randomly distributed. Therefore, the NMR spectrum denoted by $f(H)$ has the form\cite{yamada1986,kikuchi2000}
\begin{equation}
f(H) \propto \frac{H^2-H_{\rm i}^2+\omega^2/\gamma_N^2}{H_{\rm i}H^2},
\label{recta}
\end{equation}
where $\omega$ is the NMR angular frequency which is assumed to be larger than $\gamma_N H_{\rm i}$. The spectrum has two cutoff fields, $\omega/\gamma_N-H_{\rm i}$ and $\omega/\gamma_N + H_{\rm i}$, at which the spectrum has two sharp edges.
In powder samples, these sharp edges are normally smoothened because of the inhomogeneous distribution of internal fields. This effect is modeled by the Gaussian distribution function for $H_{\rm i}$. Finally, the spectra were simulated using the convolution of Eq.~(\ref{recta}) and the distribution function as\cite{kikuchi2000}
\begin{equation}
F(H) = \int_{0}^{\infty} f(H-H') g(H')\,dH',
\label{convo}
\end{equation}
where $g(H')$ is the aforementioned Gaussian distribution function. Since in the
AFM-ordered state the center of gravity of the rectangular spectra coincides
with the zero-shift position, $\omega/\gamma_N = 2.845$\,T was kept
fixed for all temperatures. As shown in Fig.~\ref{spkfit}, the simulated
spectra reproduce the edges of the experimentally-obtained broad and rectangular spectra
quite well down to 10\,K\@. This indicates that the ordered state is commensurate between 10~and 30\,K.

\begin{figure}
\includegraphics[width=3.3in]  {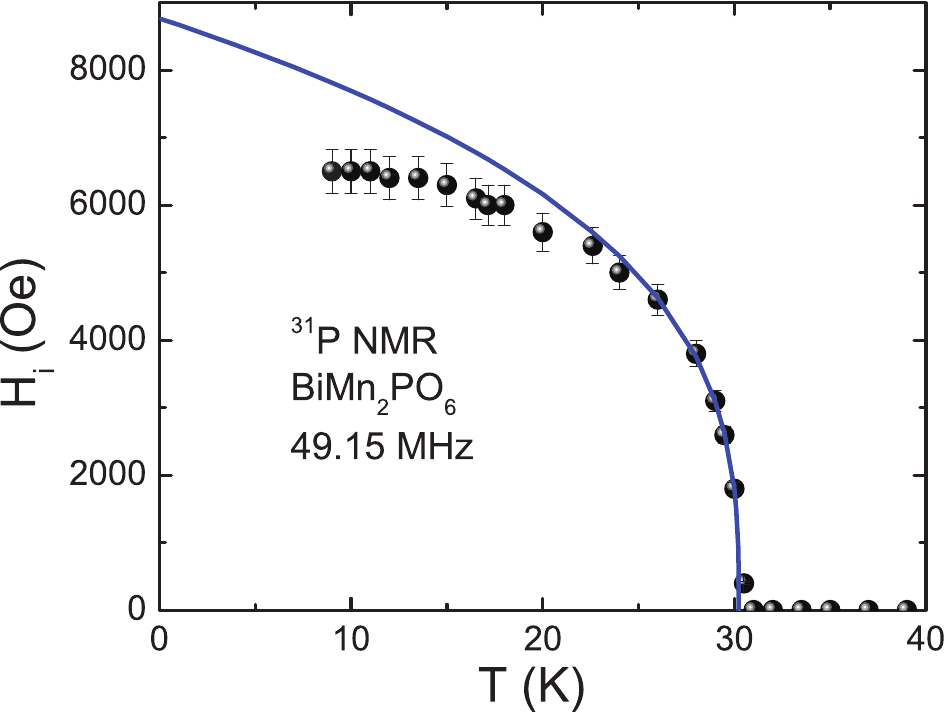}
\caption{\label{sublat} (Color online) $T$-dependence of
the internal field $H_{\rm i}$ obtained from $^{31}$P NMR spectra measured at 49.15\,MHz at $T\geq 10$\,K in the
ordered state. $H_{\rm i}$ is proportional to the Mn$^{+2}$ sublattice magnetization. The solid line is a fit of the data
above 26\,K by Eq.~(\ref{ms}) with $H_0\simeq 8760$\,Oe, $T_{\rm N} = (30 \pm 1)$\,K, and $\beta = 0.325 \pm 0.02$.}
\end{figure}

The $T$~dependence of the internal field $H_{\rm i}$ at the $^{31}$P site, which is proportional to the Mn sublattice magnetization in the ordered state, was obtained from fitting our $F(H)$ data by Eq.~(\ref{convo}) as shown in Fig.~\ref{sublat}. Below 15\,K, $H_{\rm i}(T)$ reaches saturation and remains almost constant. At higher temperatures, $H_{\rm i}(T)$ decreases as $T$ approaches $T_{\rm N}$.  In order to extract the critical exponent ($\beta$) of the order parameter
(sublattice magnetization), $H_{\rm i}$ versus $T$ was fitted by the power law
\begin{equation}
H_{\rm i}(T)=H_{0}\left(1-\frac{T}{T_{\rm N}}\right)^{\beta},
\label{ms}
\end{equation}
where $H_{0}$ is a constant. For an accurate determination of the critical exponent $\beta$,
we used data points close to $T_{\rm N}$, i.e., in the critical region. As shown in Fig.~\ref{sublat}, by fitting the data points in the temperature range 26\,K $\leq T \leq $ 30.5\,K by Eq.~(\ref{ms}) we obtained $H_{0}\simeq 8760$\,Oe, $T_{\rm N} = 30\pm 1$\,K, and $\beta = 0.325 \pm 0.02$. For comparison, we included the data points below 26\,K and arrived at the lower value of $\beta\simeq 0.27$ with $T_{\rm N}\simeq 30.13$~K\@. The critical exponent $\beta$ reflects the universality class or, equivalently, the dimensionality of the spin Hamiltonian. The expected values of $\beta$ for different universality
classes are listed in Ref.~\onlinecite{nath2009}. In BiMn$_{2}$PO$_{6}$, the fitted value of $\beta$ in the critical regime is close to the one expected for the 3D Heisenberg model, thus suggesting the 3D nature of the magnetic ordering transition at 30\,K.

\begin{figure}
\includegraphics[width=8.6cm]{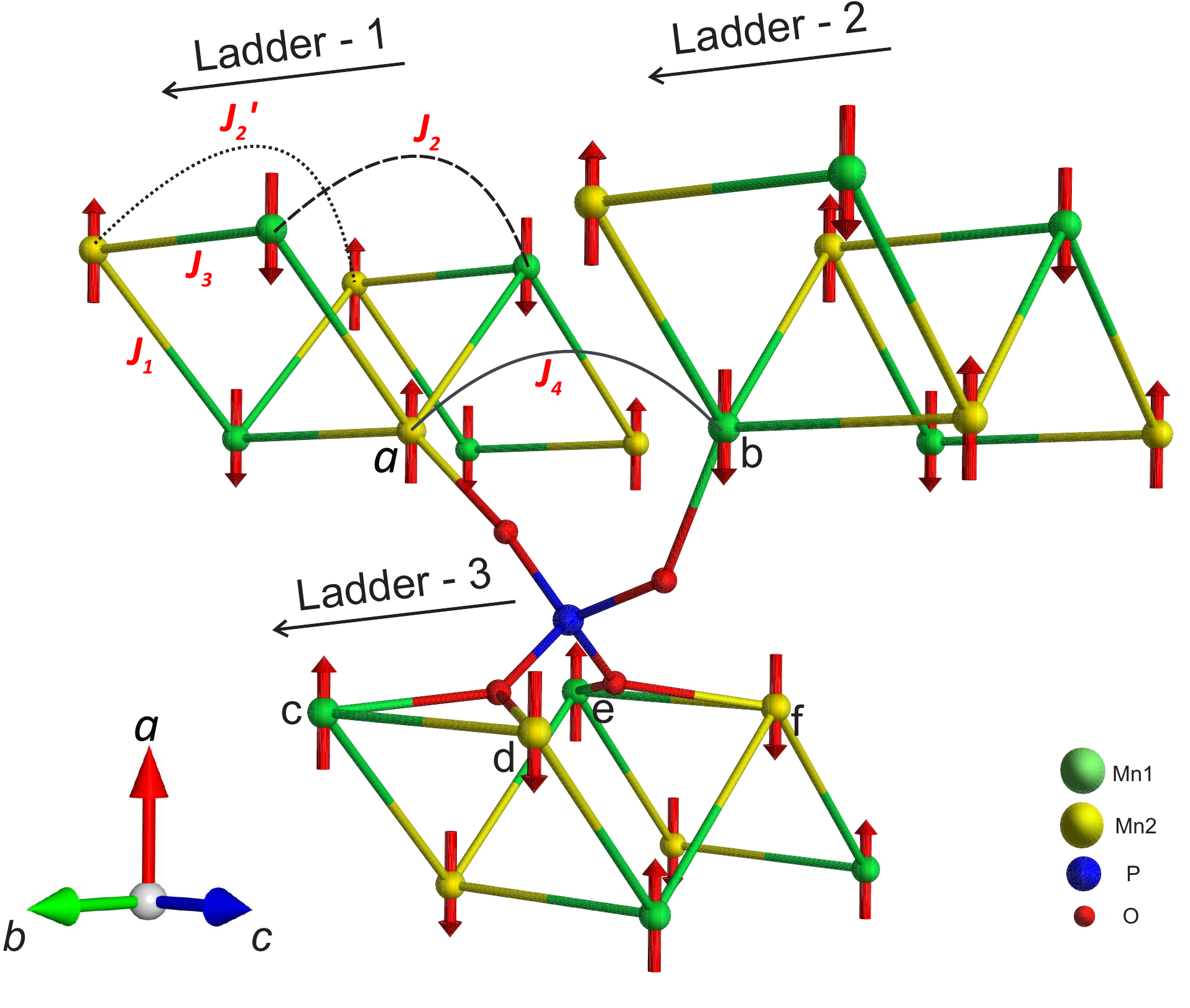}
\caption{\label{magneticstructure}
(Color online) The arrangement of $J_1-J_3$ spin ladders showing the hyperfine couplings of the P site to six neighboring Mn$^{+2}$ ions $a-f$ from three different ladders.}
\end{figure}

To understand the origin of the internal field at the P-site, we analyze the coupling of P to the Mn$^{+2}$ ions, as shown in Fig.~\ref{magneticstructure}, where each P is coupled to six Mn$^{+2}$ ions from three different ladders (site $a$ from ladder-1, site $b$ from ladder-2, and sites $c$--$f$ from ladder-3). In BiCu$_2$PO$_6$, the square-planar geometry of Cu$^{+2}$ leads to the half-filling of a single $3d$ orbital that, consequently, eliminates the hyperfine couplings to sites $a$ and $b$.\cite{bobroff2009,casola2010} By contrast, Mn$^{+2}$ features all five $3d$ orbitals half-filled, so we expect sizable hyperfine couplings to all six Mn$^{+2}$ ions around phosphorous.

As discussed above, the hyperfine field at the P-site is mainly due to the transferred hyperfine coupling, so one can understand the spin structure in the ordered state by analyzing the $^{31}$P NMR spectra. The NMR spectra were found to broaden drastically below $T_{\rm N}$ suggesting that a net field exists at the P-site due to the nearest-neighbor Mn$^{+2}$ spins. In Fig.~\ref{magneticstructure}, we show the spin configuration in the classical AFM ground state, as derived in Sec.~\ref{sec:dft} based on the GGA+$U$ results. Spins on sites $a$, $c$, and $e$ point up, whereas those on sites $b$, $d$, and $f$ point down. However, the hyperfine couplings from these spins do not cancel each other, because the orthorhombic symmetry of BiMn$_2$PO$_6$ leads to four different P--Mn distances, namely, $d_{\text{P--Mn}^a}=3.579$\,\r A, $d_{\text{P--Mn}^b}=3.321$\,\r A, $d_{\text{P--Mn}^{c,e}}=3.407$\,\r A, and $d_{\text{P--Mn}^{d,f}}=3.271$\,\r A. Therefore, a net field at the P-site is observed experimentally.

\subsection{Spin Dynamics}

In general, $\frac{1}{T_{1}T}$ is expressed in terms of the generalized susceptibility $\chi_{\rm M}(\qv,\omega)$ per mole of electronic spins as:\cite{moriya1963,mahajan1998}
\begin{equation}
\frac{1}{T_{1}T} = \frac{2\gamma_{N}^{2}k_{\rm B}}{N_{\rm A}^{2}}
\sum\limits_{\qv}\left| A(\qv)\right|
^{2}\frac{\chi^{\prime \prime}_{\rm M}(\qv,\omega)}{\omega},
\label{t1form}
\end{equation}
where the sum is over wave vectors $\qv$ within the first Brillouin zone, $A(\qv)$ is the form factor of the hyperfine interactions as a function of $\qv$, and $\chi^{\prime \prime}_{\rm M}(\qv,\omega)$ is the imaginary part of the
dynamic susceptibility at the nuclear Larmor frequency $\omega$. The uniform static molar susceptibility $\chi=\chi_{M}^{\prime}(0,0)$ corresponds to the real component $\chi_{M}^{\prime}(\qv,\omega)$ with $\qv=0$ and $\omega=0$.
In the paramagnetic regime, $1/(\chi T_{1}T)$ should remain $T$-independent.

The $1/(\chi T_{1}T)$ is plotted along the right $y$-axis in Fig.~\ref{t1}. Instead of a $T$-independent behavior, an
increase in $1/(\chi T_{1}T)$ was observed upon cooling indicating that $\sum\limits_{\qv}\mid A(\qv)\mid ^{2}\chi^{\prime \prime}_{\rm M}(\qv,\omega)$ increases more than $\chi$ does due to the growth of AFM correlations. This increase persists up to
the highest measured temperature.

At sufficiently high temperatures, $1/T_{1}$ is constant in a system with exchange-coupled local moments and can be expressed within the Gaussian approximation of the correlation function of the electronic spin as:\cite{moriya1956a}
\begin{equation}
\left(\frac{1}{T_1}\right)_{T\rightarrow\infty} =
\frac{(\gamma_{N} g\mu_{\rm B})^{2}\sqrt{2\pi}z^\prime S(S+1)}{3\,\omega_{ex}}
{\Big(\frac{A_{hf}}{z'}\Big)^{2}},
\label{t1inf}
\end{equation}
where $\omega_{ex}=\left(|J_{\rm max}|k_{\rm B}/\hbar\right)\sqrt{2zS(S+1)/3}$ is the Heisenberg exchange frequency, $z$ is the number of nearest-neighbor spins of each Mn$^{+2}$ ion, and $z^\prime$ is the number of nearest-neighbor Mn$^{+2}$ spins for a given P site. The $z^\prime$ coefficient in the numerator is due to the fact that the P site feels fluctuations arising from all
nearest-neighbor Mn$^{+2}$ spins. Using the relevant parameters, $A_{\rm hf} \simeq 7224$\,Oe/$\mu_{\rm B}$, $\gamma_N = 1.08 \times 10^8\,{\rm rad}$~s$^{-1}$\,T$^{-1}$, $z=3$, $z^\prime=6$, $g=2$, $S=\frac52$, and the high-temperature (250\,K) relaxation rate of
$\left(\frac{1}{T_1}\right)_{T\rightarrow\infty}\simeq 12\,800$\,s$^{-1}$ for the P site, the magnitude of the maximum exchange coupling constant is calculated to be $J_{\rm max}/k_{\rm B}\simeq 4.3$\,K, which is in reasonable agreement with our computed exchange couplings in Table~\ref{tab:exchanges}.

In the AFM-ordered state, $1/T_{1}$ is mainly driven by scattering of magnons off nuclear spins, leading to a power law $T$-dependence.\cite{beeman1968,belesi2006,johnston2011} For $T \gg \Delta/k_{\rm B}$, where $\Delta$ is the energy gap in the spin-wave spectrum, $1/T_{1}$ follows either a $T^{3}$ behavior due to a two-magnon Raman process or a $T^{5}$ behavior due to a three-magnon process, while for $T\ll\Delta/k_{\rm B}$, it follows an activated behavior $1/T_{1} \propto T^{2}\exp(-\Delta/k_{\rm B}T)$. As seen from Fig.~\ref{t1}, our $^{31}$P $1/T_{1}$ data below $T_{\rm N}$ follow a $T^{3}$ behavior rather than a $T^{5}$ behavior suggesting that the relaxation is mainly governed by the two-magnon Raman process. However a deviation from the power law was observed for $T \leq 10$~K which is either due to the opening of a gap $\Delta$ or due to the formation of an incommensurate or canted AFM ordering. The heat capacity data at low~$T$ argue against the spin-gap interpretation as discussed next.

\subsection{\label{Sec:SpinWaveCp} Magnetic Heat Capacity of Spin Waves}

Because the extrapolations of the $C_{\rm mag}/T^2$ and $C_{\rm mag}/T^3$ data for BiMn$_2$PO$_6$ in Fig.~\ref{Fig:BiMn2PO6_CmagOnTn} to $T=0$ appear to give nonzero intercepts, these intercepts may represent $T^2$ (2D) or  $T^3$ (3D) spin wave contributions to the heat capacity, in which case anisotropy effects are negligible in causing energy gaps in the spin-wave spectra.  Here we discuss these two potential contributions.  For 3D spin-wave propagation along the $x$, $y$ and~$z$ axes of a simple orthorhombic spin lattice, the heat capacity per mole of spins is\cite{johnston2011}
\bse
\label{Eqs:Cmag/R1}
\be
\frac{C_{\rm mag}}{R} = \left(\frac{4\pi^2V_{\rm spin}}{15\hbar^3v_xv_yv_z}\right)(k_{\rm B}T)^3,\qquad {\rm (3D)}
\label{Eq:Cmag3D}
\ee
where $V_{\rm spin}$ is the volume per spin and $v_\alpha\ (\alpha = x,y,z)$ are the respective spin-wave velocities.  For quasi-2D spin waves in the $xy$ plane, one obtains\cite{johnston2011}
\be
\frac{C_{\rm mag}}{R} = \left[\frac{6\zeta(3)A_{\rm spin}}{\pi\hbar^2v_xv_y}\right](k_{\rm B}T)^2,\qquad {\rm (2D)}
\label{Eq:Cmag2D}
\ee
\ese
where $A_{\rm spin}$ is the area per spin and $\zeta(z)$ is the Riemann zeta function.

Here we consider simple effective models of spin lattices with nearest-neighbor interactions represented by the Heisenberg Hamiltonian [Eq.~(7)].  Following Ref.~\onlinecite{johnston2011}, we take the spin wave velocities to be given by
\bse
\label{Eqs:valpha}
\bea
\hbar v_{\alpha} &=& \sqrt{6}SJ_\alpha a_\alpha \quad ({\rm 3D}, \alpha = x,\, y,\ z),\\*
\hbar v_{\alpha} &=& 2SJ_\alpha a_\alpha \qquad ({\rm 2D}, \alpha = x,\, y),
\eea
\ese
where $a_\alpha$ are the lattice parameters in the $x$, $y$ and $z$ directions, respectively.  Taking the $x$, $y$ and $z$ directions to be in the directions of the orthorhombic $a$, $b$ and $c$ crystal axes, one obtains $V_{\rm spin}=abc$ in 3D and $A_{\rm spin}=ab$ in 2D\@.  Then substituting Eqs.~(\ref{Eqs:valpha}) into~(\ref{Eqs:Cmag/R1}) gives the magnetic heat capacities per mole of spins as
\bse
\label{Eqs:CmagSW1}
\bea
\frac{C_{\rm mag}}{R} &=& \beta_{\rm SW}T^3\hspace{1in} {\rm (3D)}\label{Eq:Cmag3D2}\\*
\beta_{\rm SW}&=&  \frac{4\pi^2\sqrt{6}}{15(J_x/k_{\rm B})(J_y/k_{\rm B})(J_z/k_{\rm B})}.\nonumber
\eea
and
\bea
\frac{C_{\rm mag}}{R} &=& \delta_{\rm SW}T^2\hspace{1in} {\rm (2D)}\label{Eq:Cmag2D2}\\*
\delta_{\rm SW} &=& \frac{12\zeta(3)}{\pi(J_x/k_{\rm B})(J_y/k_{\rm B})}.\nonumber
\eea
\ese

In view of the complicated set of exchange interactions in BiMn$_2$PO$_6$ revealed by the above electronic structure calculations, here we obtain effective values $J_{\rm 2D}$ and $J_{\rm 3D}$ of the exchange constant from the heat capacity data assuming 2D or 3D propagation of spin waves and compare these values to the range of $J_{ij}$ values obtained theoretically in Table~\ref{tab:exchanges}.  Thus we define
\be
J_{\rm 3D} \equiv (J_xJ_yJ_z)^{1/3},\qquad J_{\rm 2D} \equiv (J_xJ_y)^{1/2},
\ee
and the coefficients in Eqs.~(\ref{Eqs:CmagSW1}) become
\bse
\label{Eqs:betadeltaJ}
\bea
\beta_{\rm SW}&=&  \frac{4\pi^2\sqrt{6}}{15(J_{\rm 3D}/k_{\rm B})^3},\\*
\delta_{\rm SW} &=& \frac{12\zeta(3)}{\pi(J_{\rm 2D}/k_{\rm B})^2}.
\eea
\ese
Then using the values of $\beta_{\rm SW}$ and~$\delta_{\rm SW}$ from Eqs.~(\ref{Eqs:CmagSWExpt}), Eqs.~(\ref{Eqs:betadeltaJ}) give
\be
\frac{J_{\rm 3D}}{k_{\rm B}} = 10~{\rm K},\qquad \frac{J_{\rm 2D}}{k_{\rm B}} = 68~{\rm K}.
\label{Eq:J3DJ2DValues}
\ee

The first of these values is similar to the largest AFM exchange constants in Table~\ref{tab:exchanges}.  Therefore the effective exchange coupling constants in Eq.~(\ref{Eq:J3DJ2DValues}) suggest that (i)~the connectivity of the exchange interactions is effectively three-dimensional, and~(ii) there are no significant anisotropy gaps in the spin-wave spectra with values greater than roughly 1~K\@.  The inferred 3D nature of the spatial spin interactions is consistent with the microscopic analysis of these interactions from electronic structure calculations in the following section.

\subsection{Microscopic Aspects}
\label{sec:microscopic}

\begin{figure*}
\includegraphics{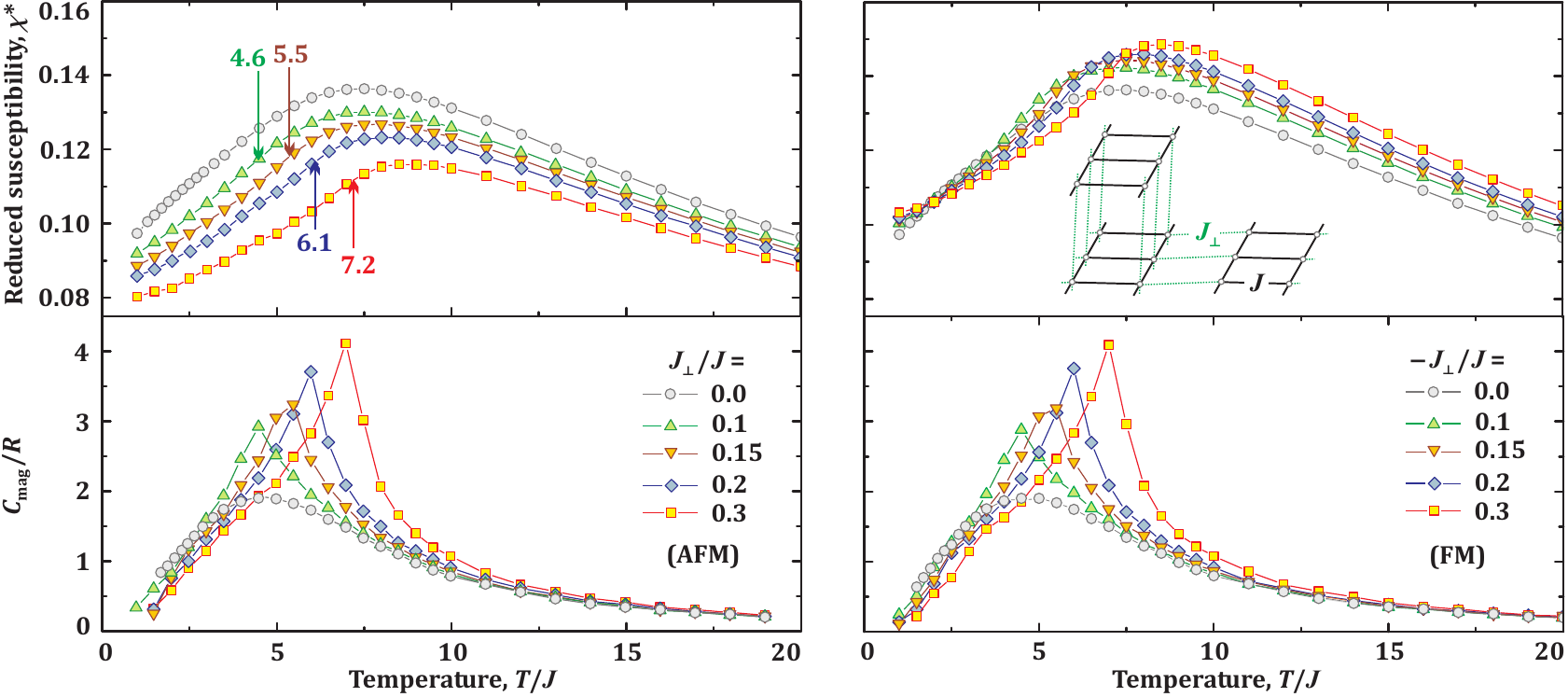}
\caption{\label{fig:simulation}
(Color online) Quantum Monte-Carlo simulations of the magnetic susceptibility ($\chi^*$, top) and specific heat ($C_{\rm mag}$, bottom) for the reference model of coupled spin-$\frac52$ ladders with the variable interladder coupling $J_{\perp}$, both AFM (left) and FM (right), see text for details. The spin lattice is depicted in the inset of the upper right panel. The arrows in the upper left panel denote N\'eel temperatures $T_{\rm N}$ determined from the peaks of the specific heat and independently verified by the scaling behavior of Binder's cumulant. Note that even a weak interladder coupling $J_{\perp}/J=0.1$ leads to a sizable $\lambda$-type anomaly superimposed on the broad maximum of $C_{\rm mag}(T)$, whereas the symmetric maximum in the magnetic susceptibility gradually transforms into an asymmetric kink.}
\end{figure*}

Here we discuss the magnetic dimensionality of BiMn$_2$PO$_6$ and the role of frustration for AFM ordering in this compound. The fact that neither the magnetic susceptibility nor the specific heat data for BiMn$_2$PO$_6$ show broad maxima at $T > T_{\rm N}$ that are typical for low-dimensional antiferromagnets indicates the spatially three-dimensional (3D) nature of the AFM Mn--Mn interactions in BiMn$_2$PO$_6$. On the other hand, individual exchange couplings reveal a pronounced 1D anisotropy. The leg and rung couplings $J_1$ and $J_3$, respectively, are at least 3 times larger than any interladder coupling (Table~\ref{tab:exchanges}).

The degree of the 1D anisotropy can be quantified by the summation of all intra- and interladder couplings per Mn site. Using the numbers in Table~\ref{tab:exchanges}, we arrive at $J_{\rm intra}=22.4 (23.0)$\,K and $J_{\rm inter}=5.3 (4.5)$\,K for Mn1(Mn2), so that $J_{\rm inter}/J_{\rm intra} = 0.22(2)$. A qualitatively different quantitative measure of the 1D anisotropy can be given by analyzing the value of the N\'eel temperature $T_{\rm N}$. A frustrated spin model with the ten couplings listed in Table~\ref{tab:exchanges} yields $T_{\rm N}\simeq 27$\,K from the classical Monte Carlo simulations in Sec.~\ref{Sec:CMC}, in excellent agreement with the experimental $T_{\rm N}\simeq 28.8$\,K obtained from the heat capacity measurements in Fig.~\ref{Fig:BiMn2PO6_CmagOnT}. Removing the frustrating interactions in Table~\ref{tab:exchanges} for which $\sisj/S^2=-1$ while keeping the quasi-1D nature of the system, we arrive at $T_{\rm N}\simeq 47$\,K from Fig.~\ref{fig:fit}. On the other hand, MFT predicts from Eq.~(\ref{Eq:TNcalc2}) that $T_{\rm N} = 68$\,K without the frustrating interactions. We conclude that frustration for AFM ordering reduces $T_{\rm N}$ by $\sim 29$\% of the initial 68~K value, and the 1D spatial anisotropy of the spin interactions reduces $T_{\rm N}$ by another $\sim 31$\%. Although qualitatively different from the simple summation of the intra- and interladder couplings, the reduction in $T_{\rm N}$ is, surprisingly, of the same scale as the ratio $J_{\rm intra}/J_{\rm inter}\simeq 0.22$.

Despite the pronounced 1D anisotropy of Mn--Mn exchange interactions, no broad maxima, which are typical signatures of the 1D physics, are seen in the magnetic susceptibility~$\chi$ and magnetic heat capacity $C_{\rm mag}$ versus~$T$ of BiMn$_2$PO$_6$ in Figs.~\ref{chi} and~\ref{cp}, respectively, as noted above. To clarify the reason for this difference in observed behaviors from the expectation for a 1D spin lattice, we consider a simplified reference model, where spin-$\frac52$ ladders with equal leg and rung couplings $J$ are connected by interladder couplings $J_{\perp}$ forming a 3D network with $z=3$ interladder couplings per site as shown schematically in the inset of the upper right panel of Fig.~\ref{fig:simulation}. This is a nonfrustrated bipartite spin lattice with only nearest-neighbor interactions. For the quantum Monte-Carlo simulations of $\chi$ and $C_{\rm mag}$ we used finite lattices with up to $24\times12\times12$ spins and periodic boundary conditions. Both FM and AFM $J_{\perp}$ were considered as shown in Fig.~\ref{fig:simulation}.  At $J_{\perp}/J=0.1$, we find from the peak in the calculated $C_{\rm mag}(T)$ that $T_{\rm N}/J\simeq 4.6$, where the long-range AFM ordering manifests itself by a large $\lambda$-type anomaly superimposed on the initial broad maximum for $J_{\perp}/J=0$ related to the 1D short-range AFM order.  The $T_{\rm N}$ is seen to increase with increasing $J_{\perp}/J$ for both FM and AFM interladder interactions.  

The broad maximum in the magnetic susceptibility versus temperature seen for $J_\perp=0$ in the lower panels of Fig.~\ref{fig:simulation} is typical of a 1D AFM spin system.  Since the uniform magnetic field needed to measure the magnetic susceptibility does not directly couple to the order parameter of an AFM, which is the staggered magnetization, the long-range ordering transition with nonzero $J_\perp$ is manifested as a maximum in the slope $d(\chi T)/dT\simeq C_{\rm mag}$ versus temperature in the lower panels of Fig.~\ref{fig:simulation} instead of a peak in $\chi(T)$.\cite{fisher1962}  Similar behaviors of $\chi(T)$ and $C_{\rm mag}(T)$ versus interlayer coupling for stacked 2D square lattices of spins $S = 5/2$ were found previously from classical Monte-Carlo simulations.\cite{johnston2011}  Thus the lack of a broad maximum in $\chi(T)$ above $T_{\rm N}$ in Fig.~\ref{chi} indicates that BiMn$_2$PO$_6$ is not a low-dimensional spin system even though spatial anisotropy in the exchange interactions is present, and must therefore be considered to be a spatially anisotropic 3D spin system.

The proclivity of spin-$\frac52$ ladders for long-range order with weak coupling between the ladders as in Fig.~\ref{fig:simulation} is rooted in the very small spin gap of an individual isolated ladder $\Delta\simeq 0.01J$.\cite{ramos2014} A two-leg spin-$\frac12$ ladder features a much larger gap $\Delta\simeq 0.5J$ that impedes or even fully eliminates long-range order when interladder couplings are weak. Thus weakly coupled spin-$\frac12$ ladders are likely to show signatures of the 1D short-range order in thermodynamic properties due to suppression of $T_{\rm N}$ via the spin gap, while spin-$\frac52$ ladders with similar interladder couplings are not as susceptible to this effect.  The lack of a broad maximum in $\chi(T)$ at $T>T_{\rm N}$ in BiMn$_2$PO$_6$ corresponds to a large 3D-like ratio $J_{\perp}/J\sim 1$, where the signature of short-range AFM ordering in $\chi(T)$ arising from low spin-lattice dimensionality at temperatures $T > T_{\rm N}$ is no longer present. 

Despite the largely 3D spatial distribution of the exchange interactions and the large spin $S=5/2$ of the Mn$^{+2}$ cations, BiMn$_2$PO$_6$ is by no means a classical antiferromagnet. As discussed above, the strong suppression of the observed $T_{\rm N}\simeq 30$\,K below the value of 68~K predicted by MFT in the absence of frustrating interactions is due to the combined effects of spatial 1D anisotropy of the spin interactions, finite spin and frustration of the spin lattice for AFM ordering. The influence of the latter three effects on the AFM structure below $T_{\rm N}$ remains an open problem that should be addressed in future studies.  In particular, we predict from our electronic structure calculations that the magnetic order below $T_{\rm N}$ is collinear and commensurate, with the propagation vector $\mathbf k=0$ and AFM order both within and between the spin ladders (Table~\ref{tab:exchanges}). Experimentally, we additionally observe subtle changes below 10\,K, tentatively attributed to a spin reorientation transition, that are not accounted for by our current microscopic model which is restricted to the purely Heisenberg Hamiltonian [Eq.~(\ref{eq:ham})]. The potential second magnetic transition at 10~K requires a more detailed investigation with neutron scattering. This transition may reflect weak effects not considered here deriving from magnetic single-ion anisotropy and/or the unusual strong lattice softening that could cause significant temperature dependences of the various Mn--Mn exchange couplings in the system. 

\section{Summary and Outlook}

Both BiMn$_2$PO$_6$ and the nonmagnetic reference compound BiZn$_2$PO$_6$ show very strong lattice {\it softening} on cooling below 200\,K, where the Debye temperature $\Theta_{\rm D}$ decreases from $\sim600$~K at room temperature to $\sim 300$\,K at low temperatures.  Most solids show much smaller variations in $\Theta_{\rm D}$ on cooling due to differences between the actual phonon density of states and that assumed in the Debye theory,\cite{Gopal1966} so the factor of two decrease in $\Theta_{\rm D}$ is very unusual. The lattice properties of these compounds certainly deserve additional investigation.

BiMn$_2$PO$_6$ is an AFM compound with a 3D topology of magnetic interactions, a significant 1D anisotropy, and a moderate frustration of interladder couplings for long-range AFM order. It develops long-range magnetic order below $T_{\rm N}\simeq 30$\,K and additionally shows a second magnetic transition around 10\,K\@.  Thermodynamic and NMR measurements suggest a commensurate magnetic order between 10\,K and 30\,K, whereas the magnetic order below 10\,K may be more complex and likely involves spin canting or incommensurate modulation.  The low-$T$ heat capacity data indicate that any energy gap in the spin-wave spectrum is $\lesssim 1$~K\@.  Microscopically, magnetic frustration, the finite spin and the 1D spatial anisotropy of the spin interactions lead to a large factor of two suppression of $T_{\rm N}$, but these features have no visible effect on the ordering pattern, at least on the classical level of the Heisenberg model with only isotropic exchange couplings that we investigated. These couplings stabilize a simple collinear N\'eel-type AFM order.

Neutron scattering experiments are particularly well suited to investigate the nature of the ordered AFM state between 10 and 30\,K as well as the AFM state below 10\,K\@. We expect that the commensurate AFM structure depicted in Fig.~\ref{structure} will be observed between 10\,K and 30\,K, whereas a more complex ordering pattern will be seen at lower temperatures. The magnetic frustration and spatial anisotropy of the exchange interactions are central to many transition-metal oxides of current interest. Additionally, quantum fluctuations associated with the finite spin and frustration for AFM ordering would be expected to suppress the zero-temperature ordered (saturation) moment from the nominal value $\mu_{\rm sat} = gS\mu_{\rm B} = 5\,\mu_{\rm B}$.\cite{johnston2011}  Delineating the role of these effects is important for building microscopic theory of complex magnetic materials.

\begin{acknowledgments}

RN and KMR acknowledge financial support from DST India.  The research at Ames Laboratory was supported by the U.S. Department of Energy, Office of Basic Energy Sciences, Division of Materials Sciences and Engineering.  Ames Laboratory is operated for the U.S. Department of Energy by Iowa State University under Contract No.~DE-AC02-07CH11358.  AT was funded by the European Union under Mobilitas grant MTT77.

\end{acknowledgments}

\end{document}